\documentclass[aps,pre,twocolumn,showpacs,superscriptaddress]{revtex4-1}
\usepackage{graphicx}
\usepackage{amsmath}
\usepackage[dvipdfm,colorlinks=true, citecolor=blue, urlcolor=blue, linkcolor=blue ]{hyperref}% add hypertext capabilities
\hypersetup{breaklinks=true}
\begin{document}
\title{Canonical versus noncanonical equilibration dynamics of open quantum systems}
\author{Chun-Jie Yang}
\affiliation{Center for Interdisciplinary Studies, School of Physical Science and Technology, Lanzhou University,
Lanzhou 730000, China}
\affiliation{Qian Xuesen Laboratory of Space Technology, China Academy of Space Technology in China Aerospace Science and Technology Corporation, Beijing 100094, China}
\author{Jun-Hong An}\email{anjhong@lzu.edu.cn}
\affiliation{Center for Interdisciplinary Studies, School of Physical Science and Technology, Lanzhou University,
Lanzhou 730000, China}
\affiliation{Centre for Quantum Technologies, National University of Singapore, 3 Science Drive 2, Singapore 117543, Singapore}
\author{Hong-Gang Luo}
\affiliation{Center for Interdisciplinary Studies, School of Physical Science and Technology, Lanzhou University,
Lanzhou 730000, China}
\affiliation{Beijing Computational Science Research Center, Beijing 100084, China}
\author{Yading Li}\email{liyd@ieee.org}
\affiliation{Qian Xuesen Laboratory of Space Technology, China Academy of Space Technology in China Aerospace Science and Technology Corporation, Beijing 100094, China}
\author{C. H. Oh}\email{phyohch@nus.edu.sg}
\affiliation{Centre for Quantum Technologies, National University of Singapore, 3 Science Drive 2, Singapore 117543, Singapore}

\begin{abstract}
In statistical mechanics, any quantum system in equilibrium with its weakly coupled reservoir is described by a canonical state at the same temperature as the reservoir. Here, by studying the equilibration dynamics of a harmonic oscillator interacting with a reservoir, we evaluate microscopically the condition under which the equilibration to a canonical state is valid. It is revealed that the non-Markovian effect and the availability of a stationary state of the total system play a profound role in the equilibration. In the Markovian limit, the conventional canonical state can be recovered. In the non-Markovian regime, when the stationary state is absent, the system equilibrates to a generalized canonical state at an effective temperature; whenever the stationary state is present, the equilibrium state of the system cannot be described by any canonical state anymore. Our finding of the physical condition on such noncanonical equilibration might have significant impact on statistical physics. A physical scheme based on circuit QED is proposed to test our results.
\end{abstract}
\maketitle

\section{Introduction}
Recently, how quantum systems, open \cite{Tasaki1998,Popescu2006,Linden2009,Goldstein2006,Reimann2010,Lychkovskiy2010,Lee2012,Genway2013} or closed \cite{MRigol2008,Ponomarev2011,Banuls2011,Polkovnikov2011}, thermalize microscopically from a certain initial state to a canonical state has attracted much attention. In statistical mechanics, based on some basic postulates, such as the equal \textit{a priori} probability postulate or the assumption of ergodicity, it can be proven kinematically that any quantum system in equilibrium with its interacting reservoir is described by a canonical state at the same temperature as the reservoir \cite{Landau1958}. However, these basic postulates rely on the subjective lack of knowledge of the system. It was revealed that they could be abandoned by examining the entanglement induced by the interaction between the system and the reservoir, by which the canonical state can be achieved without referring to ensembles or time averages \cite{Popescu2006}.

Another way to relax these assumptions in statistical mechanics from the perspective of the microscopic theory is to study the dynamics of open systems \cite{EitanGeva2000,Esposito2003,Miyashita2008,Subasi2012,Pegel2013,Adam2013}. It is expected that the equilibrium behavior of any (many-body) system in statistical mechanics should be the steady-state limit of the nonequilibrium dynamics. In this way, the study of the equilibration dynamics of an open system can give a bridge between an arbitrary initial state of the system and its statistical equilibrium state. It has been proven that \cite{EitanGeva2000,Subasi2012}, in the limit of the vanishing coupling and the memoryless correlation function of the reservoir, the steady state of the interested quantum system is a canonical state. However, in real physical situations, e.g., photonic crystal \cite{Hoeppe2012} and semiconductor phonon \cite{Tahara2011} environments, low-dimensional solid-state system \cite{Galland2008}, and even purely optical systems \cite{Liu2011,Madsen2011}, these conditions are generally invalid and the strong memory effect of the reservoir may even cause the anomalous decoherence behavior, i.e., the halting of decoherence with the increase of the coupling between the system and the reservoir \cite{An2013,Zhang2012,Liu2013}. Thus it is desirable to know how and when a nonequilibrium dynamics results in a relaxation of the system to the equilibrium states universally with respect to widely differing initial conditions and compatible to statistical mechanics.

In this work, we are interested in two questions. (1) What state does the system dynamically evolve to? (2) What is the physical condition under which the canonical state in statistical mechanics is valid? To answer these questions, we study the exact equilibration dynamics of a harmonic oscillator interacting with a thermal reservoir. A generalized canonical state with an effective temperature, which reduces to the conventional one in the vanishing limit of the coupling strength, and its validity condition are obtained microscopically. The mechanism of its breakdown is due to the formation of a stationary state of the whole system \cite{Miyamoto2005,Dreisow2008}. Our results reveal the complete breakdown of the equilibrium statistical mechanics when the stationary state is present. The possible testification of our prediction in a coupled cavity array system in circuit QED is also proposed.

Our paper is organized as follows. In Sec. \ref{Model}, we show the model of a harmonic oscillator interacting with a bosonic reservoir at finite temperature and its exact master equation derived by Feynman-Vernon's influence functional method. In Sec. \ref{thermd}, the canonical and noncanonical thermalization dynamics of the harmonic-oscillator system is revealed. In Sec. \ref{phyrezlat}, we propose an experimentally realizable scheme to test our result on noncanonical thermalization. Finally, a summary is given in Sec. \ref{Sum}.

\section{Model and exact dynamics}\label{Model}
We study a harmonic oscillator coupled to a reservoir. The Hamiltonian of the total system is $\hat{H}=\hat{H}_\text{s}+\hat{H}_\text{r}+\hat{H}_\text{i}$ with ($\hbar=1$)
\begin{eqnarray}
\hat{H}_\text{s}= \omega _{0}\hat{a}^{\dag }\hat{a},\hat{H}_\text{r}=\sum_{k} \omega _{k}\hat{b}_{k}^{\dag}\hat{b}_{k}, \hat{H}_\text{i}=\sum_{k}g_{k}(\hat{a}\hat{b}_{k}^{\dag }+\text{h.c.}), \label{Hamiltonian}
\end{eqnarray}
where $\hat{a}$ and $\hat{b}_{k}$ ($\hat{a}^{\dag}$ and $\hat{b}^{\dag}_{k}$) are, respectively, the annihilation (creation) operators of the oscillator with frequency $\omega_0$ and the $k$th reservoir mode with frequency $\omega_k$, and $g_{k}$ is the coupling strength. $g_k$ can be characterized by the spectral density
\begin{equation}
J(\omega )\equiv\sum_{k}\left\vert g_{k}\right\vert^{2}\delta (\omega -\omega _{k})=\eta \omega ( \omega/\omega _{c})^{s-1} e^{-\omega/\omega_{c}},\label{spectr}
\end{equation}
where $\omega_{c}$ is a cutoff frequency, and $\eta$ is a dimensionless coupling constant. The reservoir is classified as Ohmic if $s = 1$, sub-Ohmic if $0 < s < 1$, and super-Ohmic if $s > 1$ \cite{Leggett1987}. Equation (\ref{Hamiltonian}) is based on the rotating wave approximation, under which the energy-nonconserved terms $\hat{a}\hat{b}_k$ and $\hat{a}^\dag\hat{b}_k^\dag$ have been discarded. Note that this reduction is more than a simple approximation to the quantum Brownian motion \cite{Leggett1983,Hu1992} since our model describes an interaction conserving the total number of excitations $\hat{N}_\text{tot}\equiv \hat{a}^\dag\hat{a}+\sum_k\hat{b}^\dag_k\hat{b}_k$, a symmetry that can be imposed by nature right from the start to describe relevant experiments. Explicitly, our system is highly pertinent to quantum-optical setting where the system oscillator can describe the quantized fields in cavity \cite{Zhou2012} or in circuit \cite{Fink2008} QED, mechanical oscillators in optomechanics \cite{Kippenberg2007}, and atomic ensemble in the large-$N$ limit \cite{Hammerer2009}.

Setting $\rho _\text{tot}(0)=\rho(0)\otimes \frac{e^{-\beta \hat{H}_\text{r}}}{\text{Tr}_\text{r}e^{-\beta \hat{H}_\text{r}}}$ with $\beta=(k_BT)^{-1}$,
we can derive the exact master equation by Feynman-Vernon's influence functional method \cite{Leggett1983,Feynman1963,An2007,Xiong2010} % in the coherent-state %representation \cite{Zhang1990},see more details in Appendix \ref{app2}
\begin{eqnarray}
\dot{\rho}(t) &=&-i\Omega (t)[\hat{a}^{\dag }\hat{a},\rho (t)]  \nonumber \\
&&+[\Gamma (t)+\frac{\Gamma ^{\beta }(t)}{2}][2\hat{a}\rho (t)\hat{a}^{\dag }-\hat{a}^{\dag
}\hat{a}\rho (t)-\rho (t)\hat{a}^{\dag }\hat{a}]  \nonumber \\
&&+\frac{\Gamma ^{\beta }(t)}{2}[2\hat{a}^{\dag }\rho (t)\hat{a}-\hat{a}\hat{a}^{\dag }\rho (t)-\rho
(t)\hat{a}\hat{a}^{\dag }], \label{ExM}
\end{eqnarray}
where $\Omega (t)=-\textrm{Im}[\dot{u}(t)/u(t)]$ is the renormalized frequency, $\Gamma (t)=-\textrm{Re}[\dot{u}(t)/u(t)]$ and $\Gamma ^{\beta }(t)=\dot{v}(t)+2v(t)\Gamma(t)$ are the dissipation and noise coefficients. $u(t)$ and $v(t)$ are determined by
\begin{eqnarray}
\dot{u}(t)+i\omega _{0}u(t)+\int_{0}^{t}dt_{1}\mu(t-t_{1})u(t_{1}) =0, \label{u} \\
v(t)=\int_{0}^{t}dt_{1}\int_{0}^{t}dt_{2}u^*(t_{1})\nu(t_{1}-t_{2})u(t_{2}), \label{v}
\end{eqnarray}
under $u(0)=1$. Physically, $u(t)$, which is temperature independent and induced purely by the quantum vacuum effect, plays a role of dissipation, while $v(t)$, which is dependent on the temperature and induced by thermal effect, plays a role of fluctuation in the dynamics. Equation (\ref{v}) can serve as a nonequilibrium version of the fluctuation-dissipation relation, which ensures the positivity of $\rho(t)$ during the evolution. The kernel functions $\mu(x)=\int_{0}^{\infty }d\omega J(\omega )e^{-i\omega x}$ and $\nu(x)=\int_{0}^{\infty }d\omega J(\omega )\bar{n}(\omega)e^{-i\omega x}$ with $\bar{n}(\omega)=\frac{1}{e^ {\beta \omega }-1}$ (see Supplemental Material \cite{supp}). Equation (\ref{ExM}) keeps the same Lindblad form as the corresponding Markovian master equation, but the coefficients are time dependent. All the non-Markovian effect from the reservoir has been incorporated into these coefficients self-consistently. It can recover the Markovian one under the conditions that the system-reservoir coupling is very weak and the characteristic time scale of the reservoir correlation function is much smaller than the typical time scale of the system. Then the second-order perturbative solutions of Eqs. (\ref{u}) and (\ref{v}) read %(see details in Appendix \ref{appd})
$u(t)=e^{-(\kappa+i\omega')t}$ and $\dot{v}(t)=2\kappa \bar{n}(\omega_0)$ with $\kappa=\pi J(\omega_0)$ and $\omega'=\omega_0+\mathcal{P}\int{J(\omega)\over \omega-\omega_0}d\omega$. Thus the coefficients reduce to $\Gamma(t)=\kappa$, $\Omega(t)=\omega'$, and $\Gamma^\beta(t)=2\kappa \bar{n}(\omega_0)$, which equal the ones in the Markovian master equation (see Supplemental Material \cite{supp}).

\section{Thermalization dynamics}\label{thermd}
With the exact master equation (\ref{ExM}) at hand, the thermalization dynamics of the system can be studied readily. According to the detailed balance condition, the steady state for positive $\Gamma(t)$ and $\Gamma^\beta(t)$ can be constructed as (see Supplemental Material \cite{supp})
\begin{equation}
\rho(\infty)=\sum_{n=0}^{\infty }\frac{%
[\Gamma^\beta(\infty )/(2\Gamma(\infty ))]^{n}}{[1+\Gamma^\beta(\infty )/(2\Gamma(\infty ))]^{n+1}}|n\rangle \langle n|,\label{rhoinf}
\end{equation}
which defines a unique canonical state for any initial state
\begin{eqnarray}
\rho _\text{con}\equiv \rho(\infty)=e^{-\beta_\text{eff}\hat{H}_\text{s}}/ \text{Tr}_\text{s}e^{-\beta_\text{eff}\hat{H}_\text{s}},~(\beta_\text{eff}=1/k_B T_\text{eff}),\label{cano}
\end{eqnarray} with an effective temperature $T_\text{eff}=\frac{\omega_0 }{ k_B}\{\ln [1+\frac{2\Gamma (\infty)}{\Gamma ^{\beta }(\infty)}]\}^{-1}$.
$T_\text{eff}$ reduces exactly to the reservoir temperature $T$ in the Markovian limit. Thus the canonical ensemble assumption in statistical mechanics, i.e., a system in contact with a reservoir is described by the canonical state with the same temperature as the reservoir, can be dynamically confirmed only under the Markovian approximation. In the non-Markovian case, the equilibrated temperature $T_\text{eff}$ shows a dramatic deviation from $T$.

%From the above analysis, we can see that $\Gamma(t)$ and $\Gamma^\beta(t)$, which can be estimated by solving Eqs. (\ref{u}) and (\ref{v}),  determines the effective temperature $T_\text{eff}$.
To get a qualitative understanding on $T_\text{eff}$, we solve $u(t)$ by Laplace transform and get $u(t)=\mathcal{L}^{-1}[{1\over s+i\omega_0+\tilde{\mu}(s)}]$ with $\tilde{\mu}(s)=\int_0^\infty{J(\omega)\over s+i\omega}$. Using Cauchy residue theorem, the inverse Laplace transform can be calculated by finding the poles of the integrand, i.e.,
\begin{equation}
y(E)\equiv\omega_0-\int_0^\infty {J(\omega)\over \omega-E}d\omega=E,~(E=is).\label{yee}
\end{equation}
Note that the roots of Eq. (\ref{yee}) match well with the energy spectrum of Eq. (\ref{Hamiltonian}) in the single-excitation
subspace (see Supplemental Material \cite{supp}). It is understandable based on the fact that the vacuum-induced dissipation $u(t)$ is dominated by the single-excitation process in Eq. (\ref{Hamiltonian}). Since $y(E)$ is a monotonically decreasing function in the region $E\in (-\infty,0)$ for Eq. (\ref{spectr}), Eq. (\ref{yee}) has only one negative root if $y(0)<0$. No further discrete root exists in the region $E\in (0,+\infty)$ because it would make $y(E)$ divergent. The obtained $u(t)$ contributed from this discrete negative root, i.e. $e^{-iEt}$, corresponds to a stationary-state solution to Eq. (\ref{u}) \cite{Miyamoto2005,Dreisow2008}, which has a vanishing $\Gamma(t)$ and means a halt of decoherence \cite{An2013,Zhang2012,Liu2013}. By this criterion, the stationary state for Eq. (\ref{spectr}) is formed when
\begin{equation}
 \omega_0-\eta\omega_c\gamma(s)\le 0, \label{cond}
\end{equation}
where $\gamma(s)$ is Euler's $\Gamma$ function. The vanishing $\Gamma(t)$ would result in the divergence of $T_\text{eff}$. Hence, we argue that the canonical equilibration would be invalid when the stationary state is formed. In this case, the dynamics will keep the superposition amplitude of the formed stationary state unchanged, which causes a notable contribution of the stationary state in the equilibrium state. It is the reason for the breakdown of the canonical equilibration.

\begin{figure}[tbp]
  \centering
  \includegraphics[width=0.48\columnwidth]{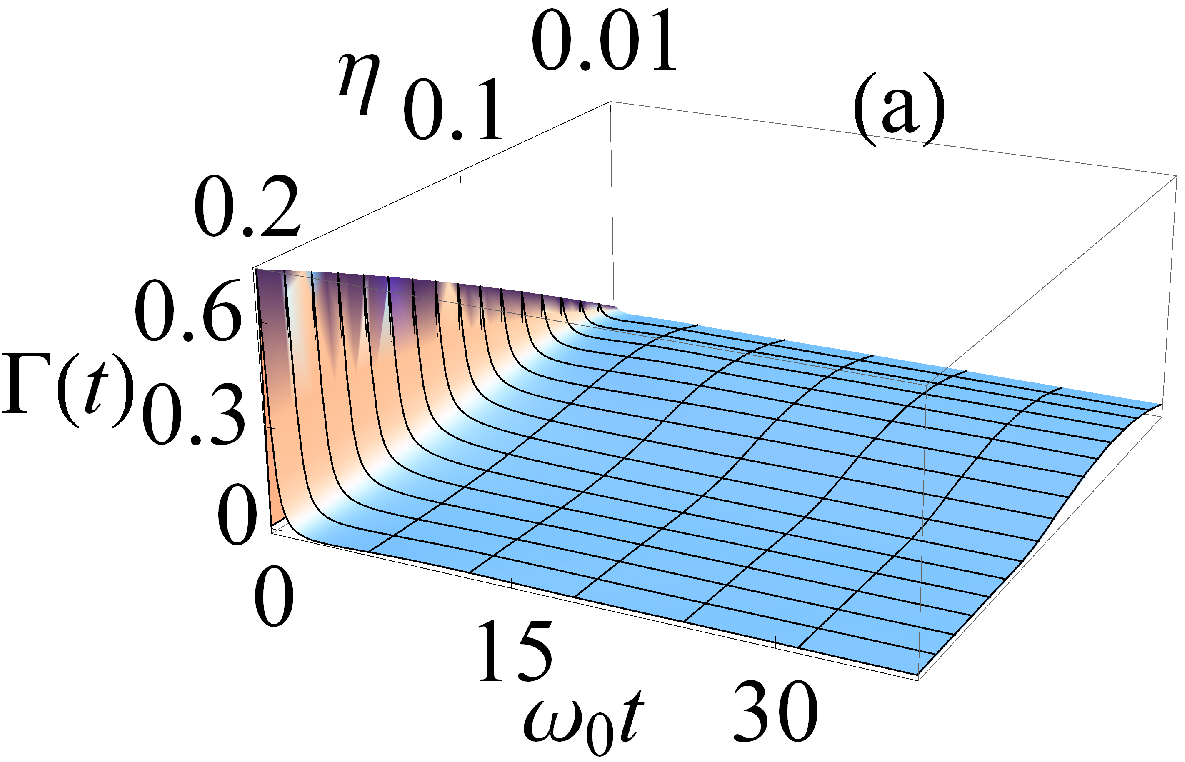}~\includegraphics[width=0.48\columnwidth]{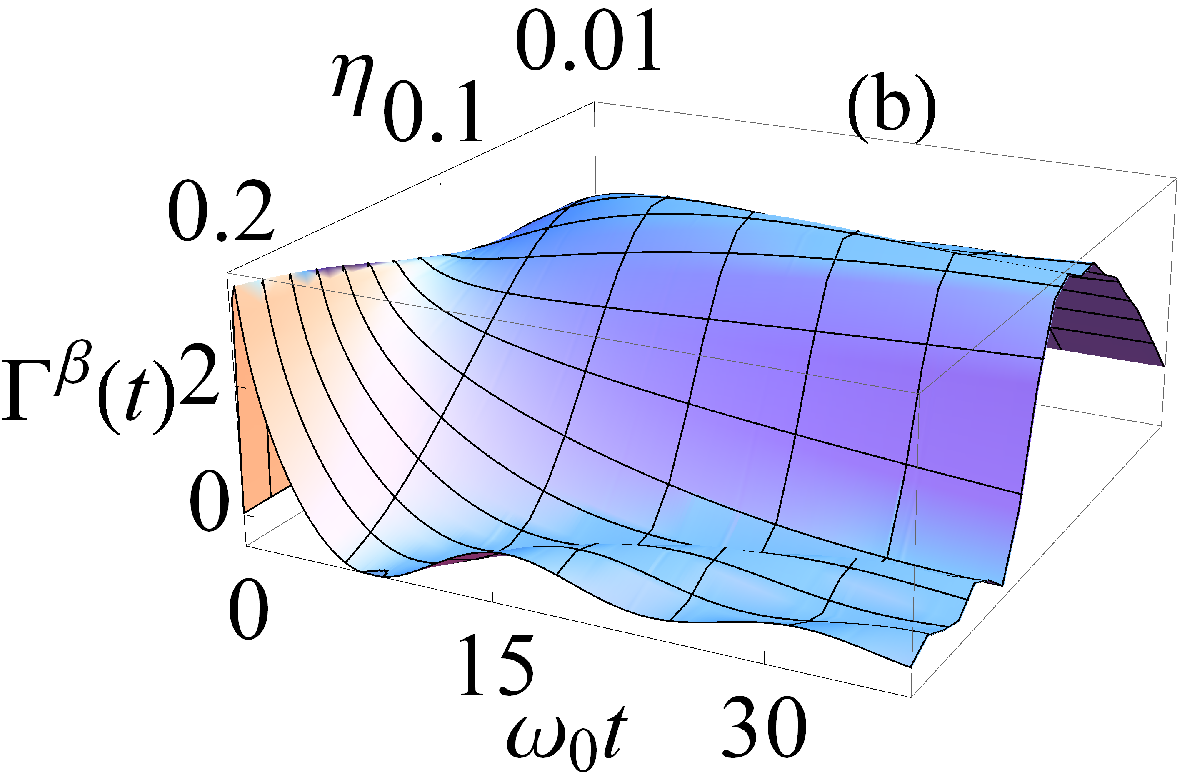}\\
  \includegraphics[width=0.96\columnwidth]{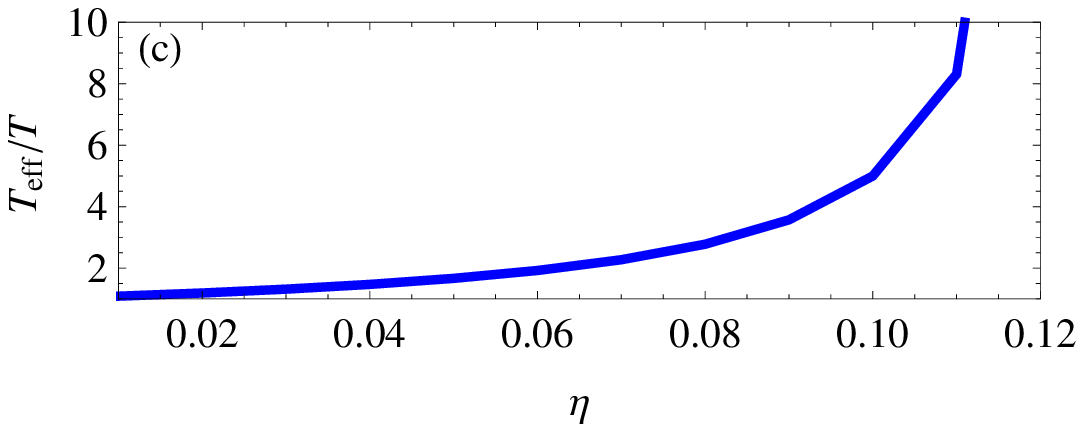}
  \caption{(Color online) $\Gamma(t)$ (a) and $\Gamma^\beta(t)$ (b) in different $\eta$. (c) $T_\text{eff}$ evaluated when $t=100/\omega_0$. $\omega_c=8.0\omega_0$, $\beta=0.1/\omega_0$, and $s=1$ have been used.}\label{difeta}
\end{figure}
\begin{figure}[h]
  \centering
  \includegraphics[width=0.48\columnwidth]{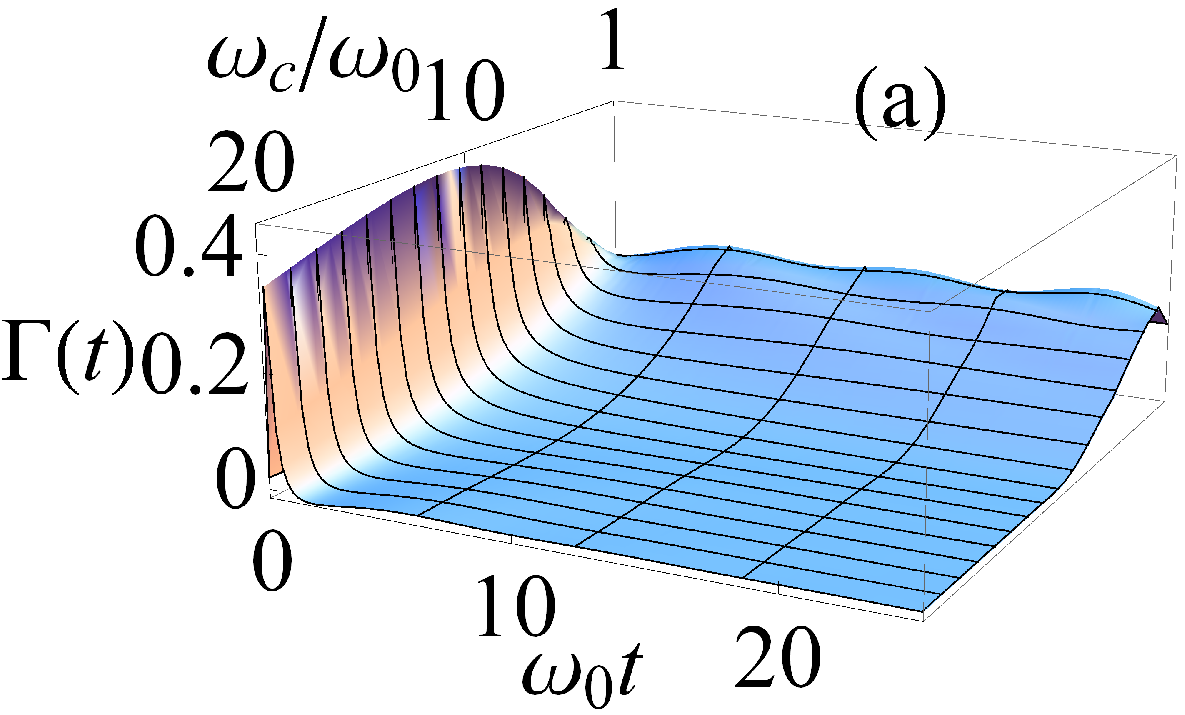}~\includegraphics[width=0.48\columnwidth]{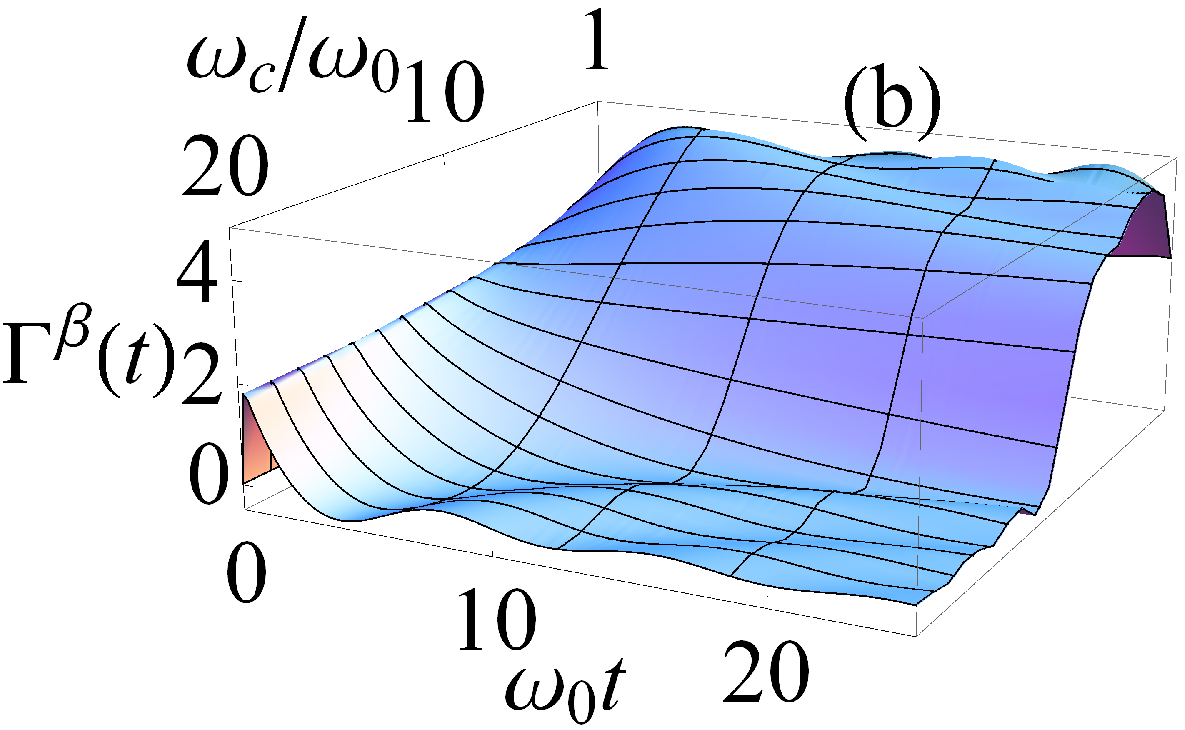}\\
  \includegraphics[width=0.96\columnwidth]{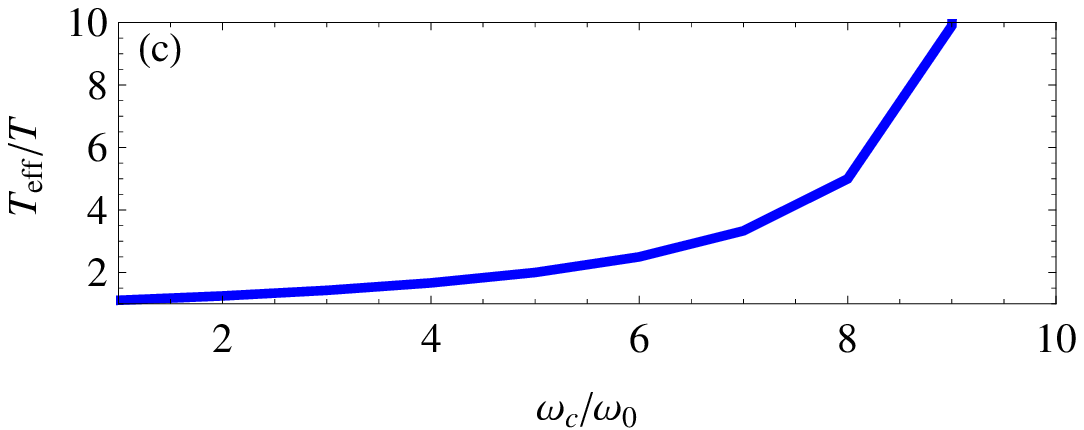}
  \caption{(Color online) $\Gamma(t)$ (a) and $\Gamma^\beta(t)$ (b) in different $\omega_c$. (c) $T_\text{eff}$ evaluated when $t=100/\omega_0$. $\eta=0.1$, $\beta=0.1/\omega_0$, and $s=1$ have been used. }\label{difomegac}
\end{figure}

To verify the validity of Eq. (\ref{cano}) and evaluate the consequence of the formed stationary state on the equilibration of the system, we present numerical simulations on the equilibration dynamics. For concreteness, we choose the Ohmic spectral density, i.e., $s=1$ in Eq. (\ref{spectr}), to do the numerics. Then we can obtain from the condition (\ref{cond}) that the stationary state is formed when $\omega_0\le \eta\omega_c$. Figures \ref{difeta} and \ref{difomegac} show the obtained $\Gamma(t)$, $\Gamma^\beta(t)$, and $T_\text{eff}$ in different parameter regimes. We can see that when the stationary state is absent, i.e., $\eta<0.125$ in Fig. \ref{difeta} and $\omega_c< 10\omega_0$ in Fig. \ref{difomegac}, both $\Gamma(t)$ and $\Gamma^\beta(t)$ tend to positive constant values asymptotically. The positivity of the two coefficients guarantees the overall thermalization of the system during the dynamics. Consequently, the system approaches a canonical state governed by Eq. (\ref{cano}), as shown in Figs. \ref{difeta}(c) and \ref{difomegac}(c), irrespective of in what state the system initially resides. Here it is qualitatively consistent with the one under the Born-Markovian approximation. However, whenever the stationary state is formed, i.e., $\eta\ge0.125$ in Fig. \ref{difeta} and $\omega_c\ge 10\omega_0$ in Fig. \ref{difomegac}, $\Gamma(t)$ and $\Gamma^\beta(t)$ possess transient negative values and approach zero asymptotically, which confirms our expectation based on the stationary-state analysis. The negative coefficients reveal a nonzero non-Markovianity \cite{Breuer}, which in turn gives an insight to the dynamical consequence of the formed stationary state (see Supplemental Material \cite{supp}). The vanishing of the two coefficients causes the dissipationless dynamics \cite{Tong2010,Zhang2012}, which is totally different from the Markovian case, and the ill definition of the canonical state (\ref{cano}). In Figs. \ref{difeta}(c) and \ref{difomegac}(c), we really see that $T_\text{eff}$ changes to be divergent when the parameters tend to the critical point forming the stationary state. Another interesting observation is that even when the equilibration to a canonical state is valid in the absence of the stationary state, $T_\text{eff}$ still shows a dramatically quantitative difference to $T$. They match well only when $\eta$ is vanishingly small [see Fig. \ref{difeta}(c)], and when $\omega_c$ is comparable with $\omega_0$ [see Fig. \ref{difomegac}(c)].

\begin{figure}[h]
  \centering
  \includegraphics[width=0.48\columnwidth]{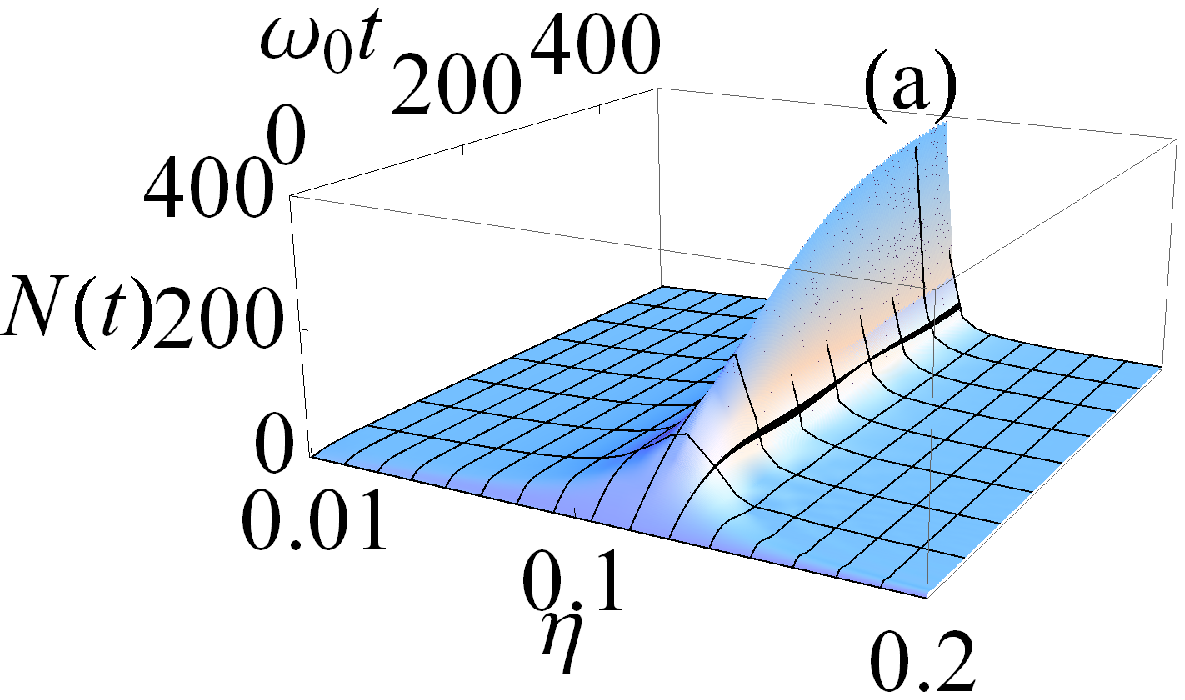}~~\includegraphics[width=0.48\columnwidth]{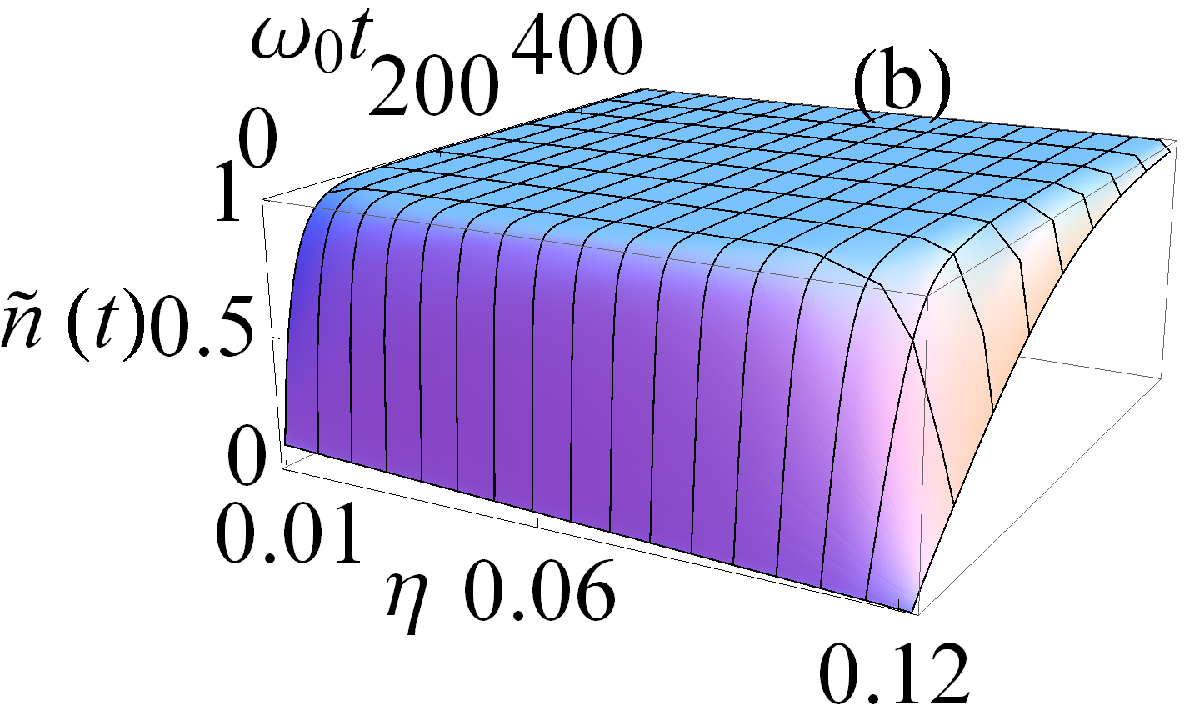}\\
  \includegraphics[width=0.96\columnwidth]{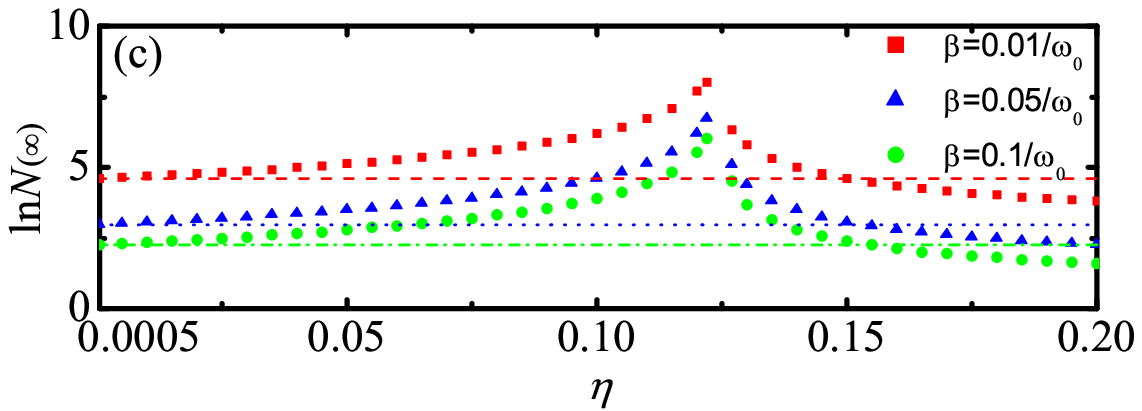}
  \caption{(Color online) $N(t)$ (a) and $\tilde{n}(t)\equiv {N(t)/ N_\text{con}}$ (b) in different $\eta$. $|\alpha_0|=1.0$ and other parameters are the same as Fig. \ref{difeta}. (c) $\ln N(\infty)$ and the corresponding Markovian values shown by dashed, dotted, and dot-dashed lines in different $T$.}\label{expectat}
\end{figure}
\begin{figure}[h]
  \centering
  \includegraphics[width=0.48\columnwidth]{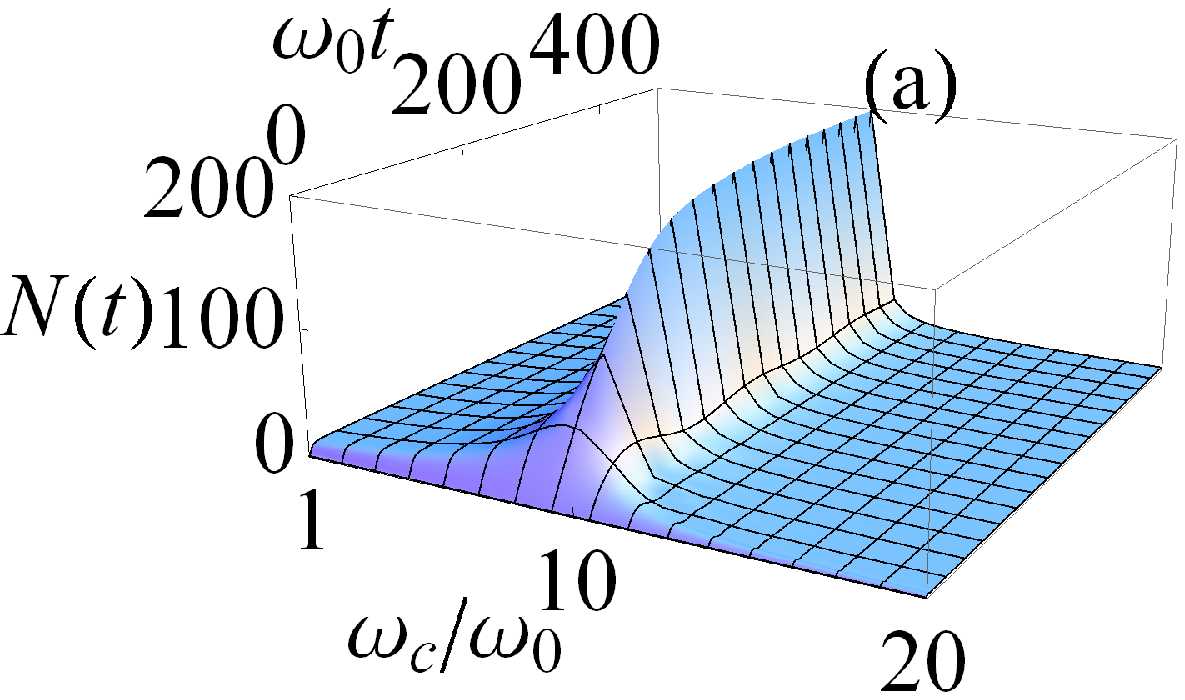}~~\includegraphics[width=0.48\columnwidth]{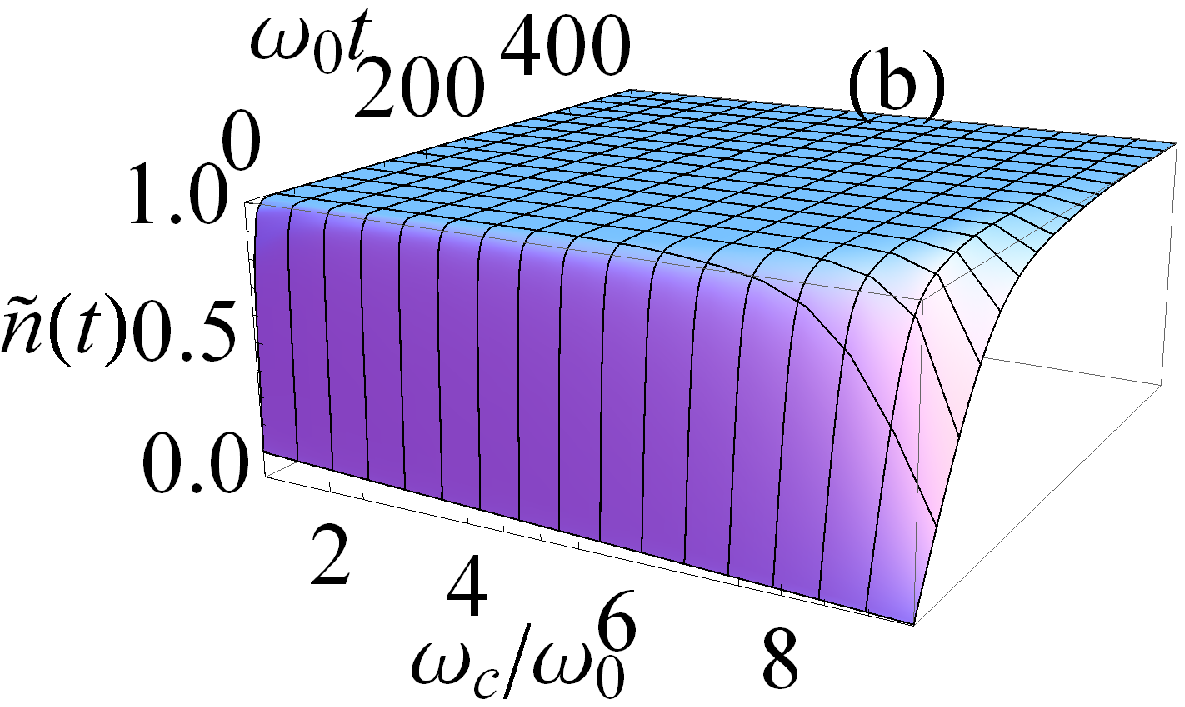}\\
  \includegraphics[width=0.96\columnwidth]{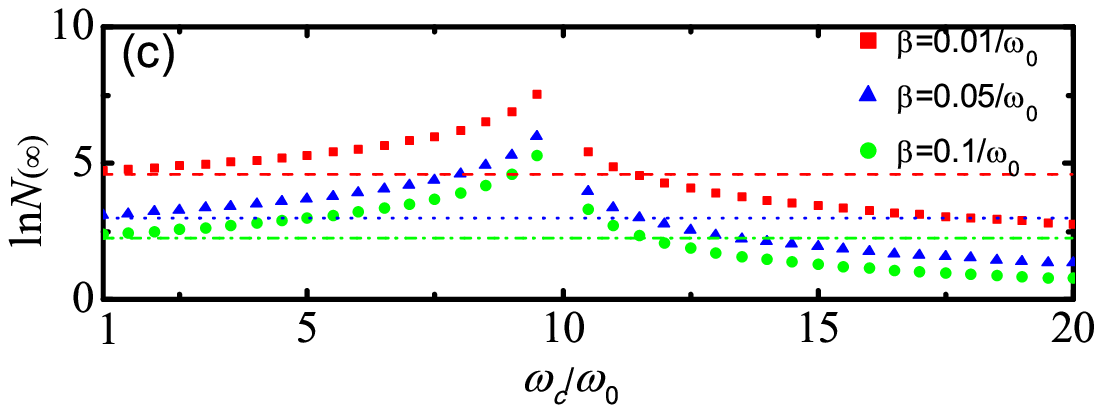}
  \caption{(Color online) $N(t)$ (a) and $\tilde{n}(t)\equiv {N(t)/ N_\text{con}}$ (b) in different $\omega_c$. $|\alpha_0|=1.0$ and other parameters are the same as Fig. \ref{difomegac}. (c) $\ln N(\infty)$ and the corresponding Markovian values shown by dashed, dotted, and dot-dashed lines in different $T$.}\label{expectaomeg}
\end{figure}
We next investigate an actual example by employing an explicit initial state of the system harmonic oscillator as $\rho(0)=e^{-|\alpha_0|^2}|\alpha_0\rangle\langle \alpha_0|$. In the framework of the influence-functional method, we can readily calculate its time evolution state and the expectation value of the bosonic number operator $N(t)=|\alpha_0u(t)|^2+v(t)$.
Figures \ref{expectat}(a) and \ref{expectaomeg}(a) show the evolution of $N(t)$ in the full parameter regime. We can see clearly that $N(t)$ behaves divergently at the critical point forming the stationary state, i.e., $\eta=0.125$ in Fig. \ref{expectat}(a) and $\omega_c=10\omega_0$ in Fig. \ref{expectaomeg}(a). We plot in Figs. \ref{expectat}(b) and \ref{expectaomeg}(b) the ratio $N(t)$ to $N_\text{con}\equiv \text{Tr}_\text{s}[\hat{a}^\dag\hat{a}\rho_\text{con}]$ in the parameter regime where the stationary state is absent. We can see that the ratio in both of the results exclusively tends to 1 in the long-time limit. It verifies the validity of Eq. (\ref{cano}) to describe the equilibrium state of the system. We show in Figs. \ref{expectat}(c) and \ref{expectaomeg}(c) the long-time values of bosonic number $N$ in different initial temperatures of the reservoir. We can see that, irrespective of the initial temperature, $N(\infty)$ diverges at the same critical point. It means that the condition for the formation of the stationary state is independent of the initial temperature of the reservoir, which confirms our analytical criterion (\ref{cond}).

\begin{figure}[h]
  \centering
  \includegraphics[width=0.49\columnwidth]{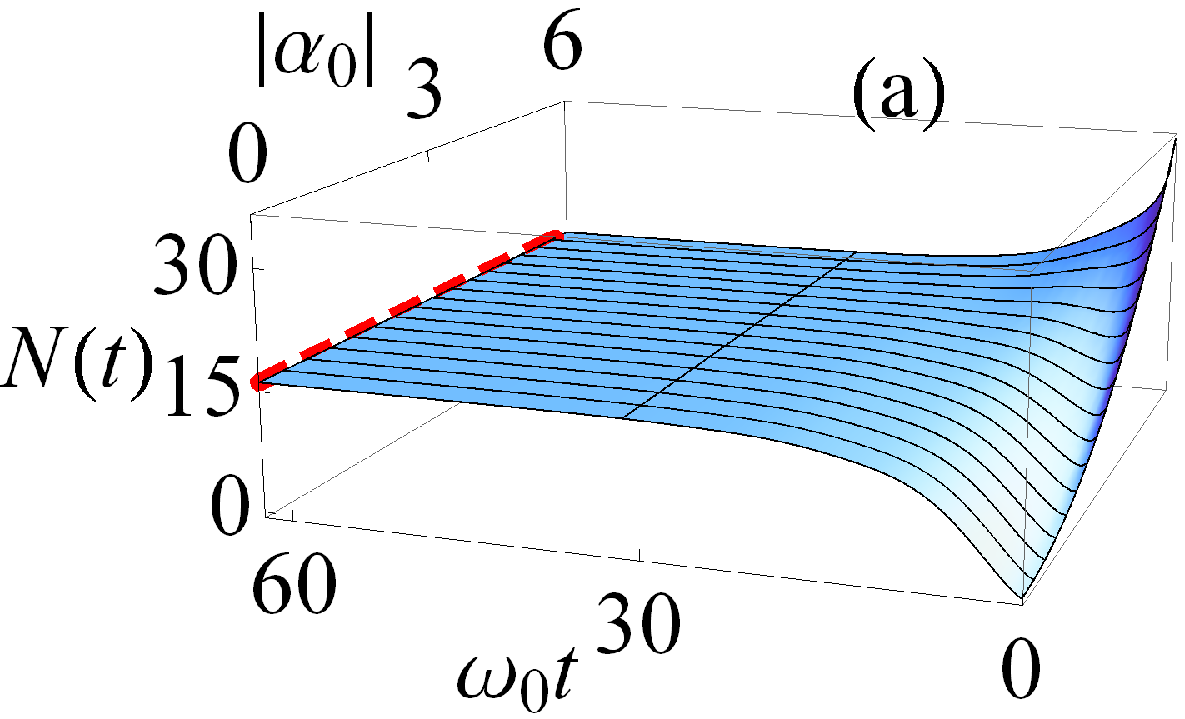}~~\includegraphics[width=0.49\columnwidth]{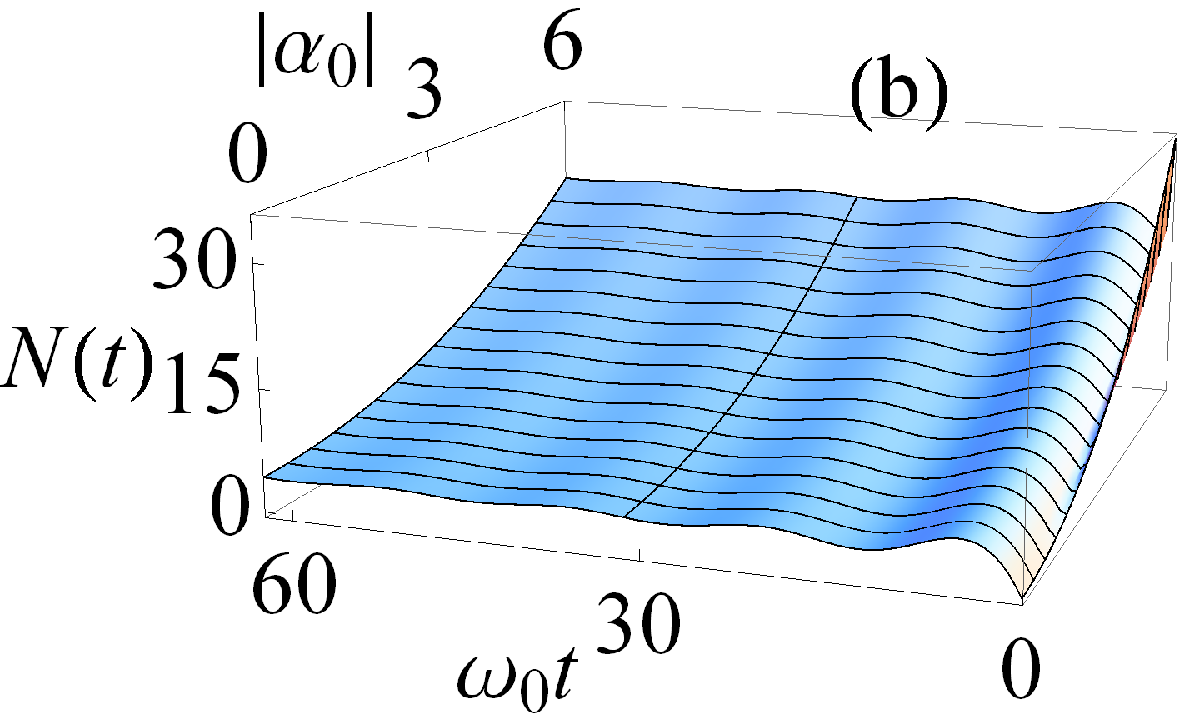}
  \caption{(Color online) $N(t)$ in different $\alpha_0$ when $\eta=0.05$ (a) and $0.2$ (b). Other parameters are the same as Fig. \ref{difeta}. The dashed line in (a) is the value evaluated from Eq. (\ref{cano}). }\label{initcc}
\end{figure}
%\begin{figure}[h]
%  \centering
 % \includegraphics[width=0.47\columnwidth]{Fig6(a).eps}~~\includegraphics[width=0.47\columnwidth]{Fig6(b).eps}
  %\caption{(Color online) $N(t)$ in different $\alpha_0$ when $\omega_c=1.0\omega_0$ (a) and $20.0\omega_0$ (b). Other parameters are the same as Fig. \ref{difomegac}. The dashed line in (a) is the value evaluated from Eq. (\ref{cano}).  }\label{initcd}
%\end{figure}
To verify the uniqueness of the equilibrium state when the stationary state is absent, we plot in Fig. \ref{initcc} the time evolution of $N(t)$ in different initial condition $\alpha_0$. Figure \ref{initcc}(a) indicates that, when the stationary state is absent, $N(t)$ tends to the unique steady value $N_\text{con}$. However, when the stationary state is formed, the steady value of $N(t)$ is not unique and dependent on the initial condition $\alpha_0$, which also proves the breakdown of the thermalization to a canonical state in this situation. This corresponds to a situation which cannot be described by the equilibrium statistical mechanics. The similar result can also be achieved in different $\omega_c$ regimes with definite $\eta$. The initial-state dependence of Fig. \ref{initcc}(b) also reveals the incorrectness of the result of Ref. \cite{Subasi2012}, which is initial-state independent, in describing our system.

\begin{figure}[tbp]
  \centering
  \includegraphics[width=0.99\columnwidth]{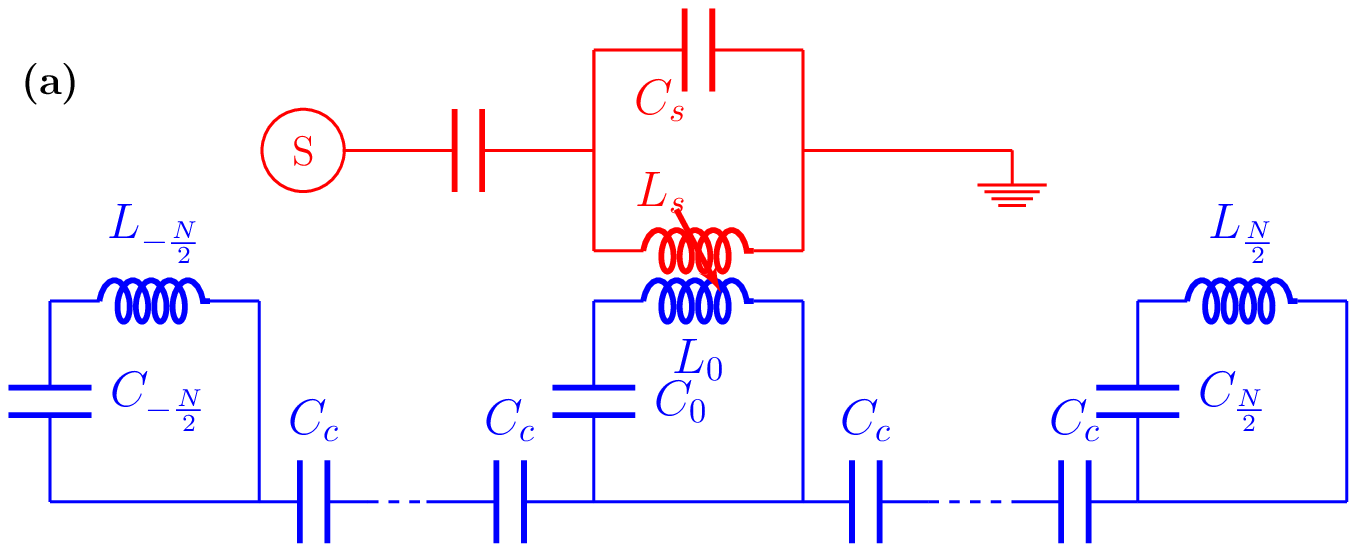}\\
  \includegraphics[width=0.49\columnwidth]{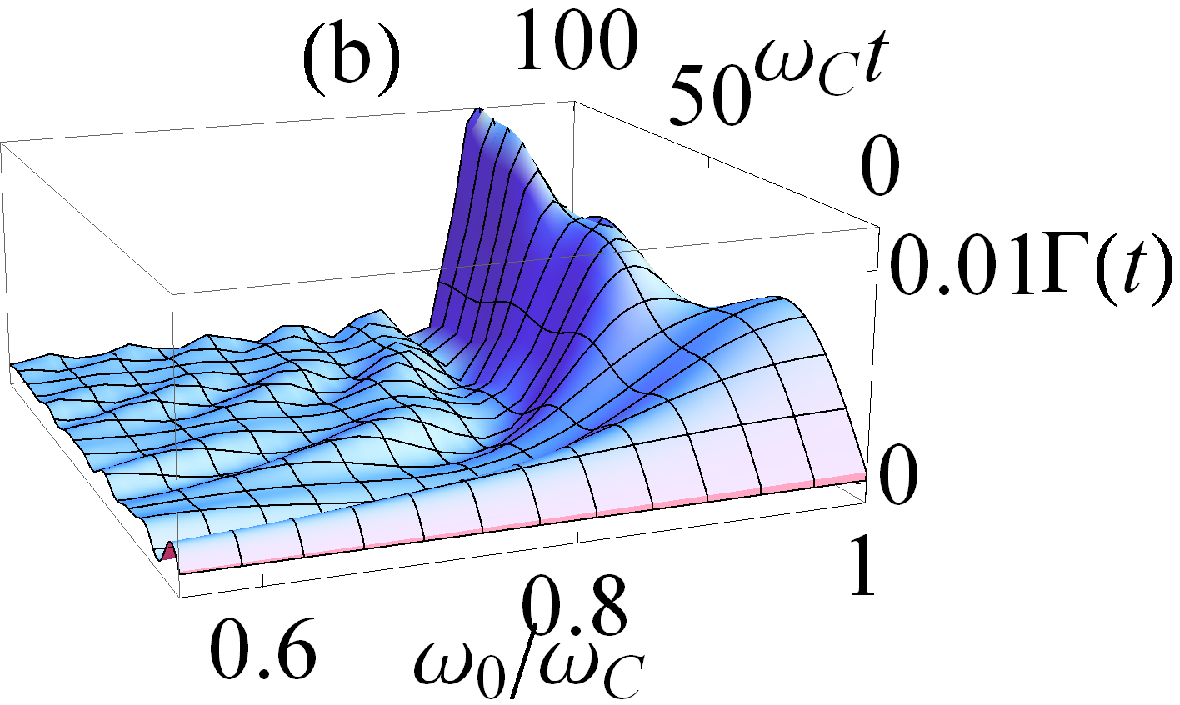}~~\includegraphics[width=0.49\columnwidth]{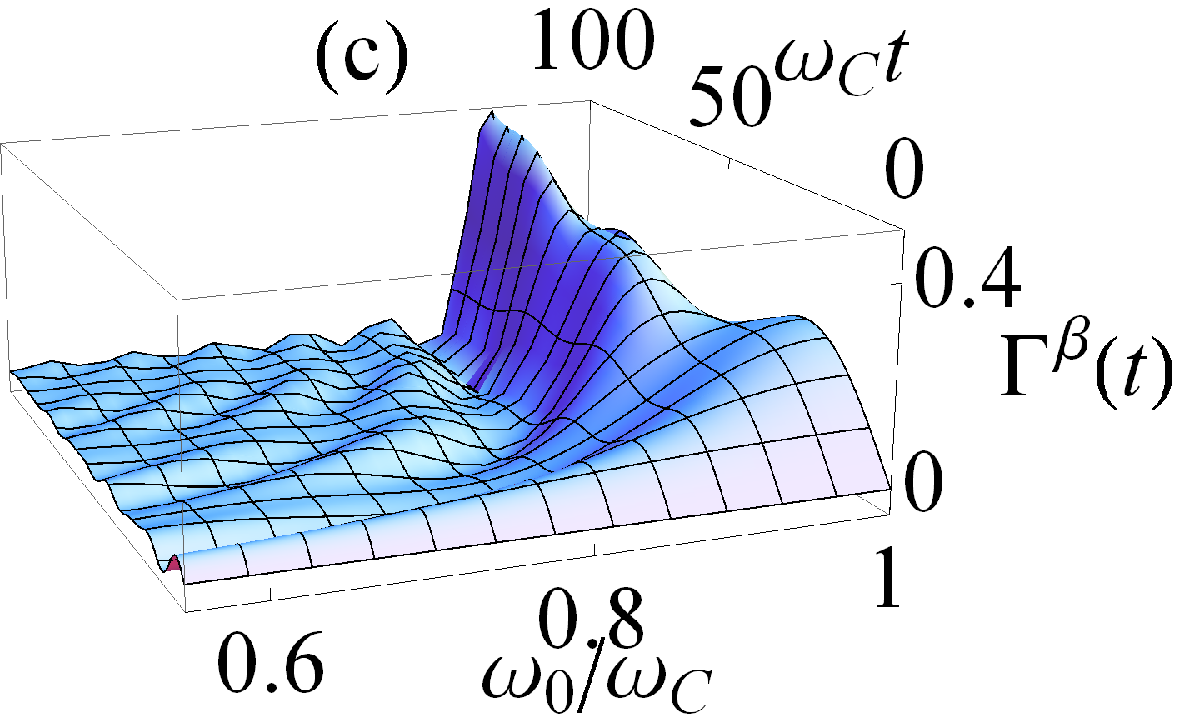}
  \caption{(Color online) (a) Scheme on a TLR (formed by $C_\text{s}$ and $L_\text{s}$) with $\omega_0={1/( 2\sqrt{L\text{s}C_\text{s}})}$ interacting with an array of TLRs (formed by $C_i$ and $L_i$) with identical $\omega_C={1/ (2\sqrt{L_{i}C_{i}})}$ coupled via $C_c$ in strength $\xi=2Z_0C_c\omega_C^2$ ($Z_0$ being the impedance). $\Gamma(t)$ (b) and $\Gamma^\beta(t)$ (c) in different $\omega_0$. $\xi=0.05\omega_C$, $g=0.02\omega_C$, $\beta=0.05/\omega_C$, and $N=200$ have been used.  }\label{scm}
\end{figure}

\section{Physical realization}\label{phyrezlat}
A promising physical system to test our prediction is an array of coupled cavities, which can now be realized experimentally in an optical-ring resonator system \cite{Hafezi2013}, in a microdisk cavity system \cite{Barclay2006}, in a photonic crystal system \cite{Hennessy2007,Notomi2008,Majumdar2012}, and synthesized in optical waveguides \cite{Verbin2013,Rechtsman2012}. In Fig. \ref{scm}, we depict a scheme realized in a circuit QED system \cite{Underwood2012,Houck2012,Egger2013}. Here, the system is realized by the transmission line resonator (TLR), and the reservoir is realized by an array of capacitance coupled TLRs, which in turn couples to the system via inductance. The reservoir is governed by $\hat{H}_\text{r}=\omega_C\sum_{j=-N/ 2}^{N/2}\hat{ b}^\dag_j \hat {b}_j+(\xi\sum_{j=-N/2}^{N/2}\hat{b}_{j+1}^\dag\hat{b}_j+\text{h.c.})$ and the interaction is $\hat{H}_\text{i}=g\hat{a}^\dag \hat{b}_0+\text{h.c.}$. $\hat{H}_\text{r}$ is recast into $\hat{H}_\text{r}=\sum_{k}\epsilon_k\hat{ b}^\dag_k \hat {b}_k$ by a Fourier transform $\hat{b}_j=\sum_k\hat{b}_ke^{ikjx_0}$ with $\epsilon_k=\omega_C+2\xi\cos kx_0$ and $x_0$ being the spatial separation of the two neighbor resonators. One can notice that the dispersion relation of the reservoirs shows finite band width $4\xi$ centered at $\omega_C$, which can induce a strong non-Markovian effect even in the weak- and intermediate-coupling regimes. The stationary state is formed if $\omega_0$ falls in the gap of the spectrum of the reservoir, i.e., $\omega_0<\omega_C-2\xi$ \cite{An2013}. We really see in Figs. \ref{scm}(b) and (c) that both $\Gamma(t)$ and $\Gamma^\beta(t)$ calculated based on an experimentally accessible parameter regime \cite{Underwood2012,Houck2012} approach zero in this situation, where the equilibration to a canonical state is expected to be invalid.

\section{Conclusion}\label{Sum}
In summary, we have studied the non-Markovian equilibration dynamics of a harmonic oscillator coupled to a reservoir. A generalized canonical state is constructed microscopically. It recovers the conventional canonical state in the Markovian approximation, but gets to be invalid whenever the stationary state is formed. Our work gives a bridge between the ensemble theory of the equilibrium statistical mechanics and the non-equilibrium dynamics, thus builds a unified framework to treat the issues in equilibrium and nonequilibrium statistical mechanics. In particular, the explicit criterion for the formation of the stationary state may supply a guideline to experiments on exploring the noncanonical equilibration of open system via engineering the specific structure of the reservoir \cite{Barreiro2011,Murch2012}.

\section*{ACKNOWLEDGEMENTS}
This work is supported by the Fundamental Research Funds for the Central Universities, by the Specialized Research Fund for the Doctoral Program of Higher Education, by the Program for NCET, by the NSF of China (Grant No. 11175072, No. 11174115, No. 11325417, and No. 61271145), and by the National Research Foundation and Ministry of Education,
Singapore (Grant No. WBS: R-710-000-008-271).

\textit{Note added}: Recently, we became aware of two related works \cite{Sun2013,Xiong2013}.

\clearpage
\noindent

\noindent

\onecolumngrid

%%%%%%%%%%%%%%%%%%%%%%%%%%%%%%%%%%%%%%
%%   Supplementary Information
%%%%%%%%%%%%%%%%%%%%%%%%%%%%%%%%%%%%%%
%\appendix
\renewcommand{\thesection}{S-\arabic{section}}
\renewcommand{\theequation}{S\arabic{equation}}
\setcounter{equation}{0}  %  this will re-count eq from 1
\renewcommand{\thefigure}{S\arabic{figure}}
\setcounter{figure}{0}  %  this will re-count eq from 1

\section*{\Large\bf Supplemental Material}
\setcounter{section}{0}
\twocolumngrid
\section{The exact decoherence dynamics}

In this section, we give the detailed derivation of exact master equation
using Feynman-Vernon's influence-functional theory \cite%
{Feynman1963,Leggett1983,Hu1992,An2007} in the coherent-state representation
\cite{Zhang1990}.

\subsection{Influence functional method}

\label{app2} The Hamiltonian of the total system reads $\hat{H}=\hat{H}_{%
\text{s}}+\hat{H}_{\text{r}}+\hat{H}_{\text{i}}$ with ($\hbar =1$)
\begin{equation}
\hat{H}_{\text{s}}=\omega _{0}\hat{a}^{\dag }\hat{a},\hat{H}_{\text{r}%
}=\sum_{k}\omega _{k}\hat{b}_{k}^{\dag }\hat{b}_{k},\hat{H}_{\text{i}%
}=\sum_{k}g_{k}(\hat{a}\hat{b}_{k}^{\dag }+\text{h.c.}),
\end{equation}
The total density matrix obeys $i\dot{\rho}_{\text{tot}}(t)$ $=[\hat{H}%
,\rho_{\text{tot}}(t)]$, we have the formal solution as
\begin{equation}
\rho _{\text{tot}}\left( t\right) =e^{-i\hat{H}t}\rho _{\text{tot}}(0)e^{i%
\hat{H}t}.  \label{fs}
\end{equation}%
In the coherent-state representation, Eq. (\ref{fs}) is expanded as
\begin{eqnarray}
&&\langle \bar{\alpha}_{f},\mathbf{\bar{z}}_{f}|\rho _{\text{tot}%
}\left(t\right) |\alpha _{f}^{\prime },\mathbf{z}_{f}\rangle =\int d\mu (%
\mathbf{z}_{i})d\mu (\alpha _{i})d\mu (\mathbf{z}_{i}^{\prime })d\mu
(\alpha_{i}^{\prime })  \notag \\
&&\times \langle \bar{\alpha}_{f},\mathbf{\bar{z}}_{f};t|\alpha _{i},\mathbf{%
z}_{i};0\rangle \langle \bar{\alpha}_{i},\mathbf{\bar{z}}_{i}|\rho _{\text{%
tot}}(0)|\alpha _{i}^{\prime },\mathbf{z}_{i}^{\prime }\rangle  \notag \\
&&\times \langle \bar{\alpha}_{i}^{\prime },\mathbf{\bar{z}}%
_{i}^{\prime};0|\alpha _{f}^{\prime },\mathbf{z}_{f};t\rangle ,  \label{tot}
\end{eqnarray}%
We assume that the initial density matrix factors into a system part and an
environment part $\rho _{\text{tot}}(0)=\rho (0)\otimes \rho _{\text{r}}(0)$%
. The dynamics of the system can be obtained by integrating over the
environmental variables,
\begin{eqnarray}
\rho (\bar{\alpha}_{f},\alpha _{f}^{\prime };t) &=&\int d\mu
(\alpha_{i})d\mu (\alpha _{i}^{\prime })\mathcal{J}(\bar{\alpha}%
_{f},\alpha_{f}^{\prime };t|\bar{\alpha}_{i}^{\prime },\alpha _{i};0)  \notag
\\
&&\times \rho (\bar{\alpha}_{i},\alpha _{i}^{\prime };0),  \label{timeevo}
\end{eqnarray}%
where $\rho (\bar{\alpha}_{f},\alpha _{f}^{\prime };t)\equiv \int d\mu (%
\mathbf{z}_{f})\langle \bar{\alpha}_{f},\mathbf{\bar{z}}_{f}|\rho _{\text{tot%
}}\left( t\right) |\alpha_{f}^{\prime },\mathbf{z}_{f}\rangle $. The
effective propagation functional of the density matrix is defined as
\begin{eqnarray}
&\mathcal{J}(\bar{\alpha}_{f},\alpha _{f}^{\prime };t|\bar{\alpha}%
_{i}^{\prime },\alpha _{i};0)=\int d\mu (\mathbf{z}_{f})d\mu (\mathbf{z}%
_{i})d\mu (\mathbf{z}_{i}^{\prime })&  \notag \\
&\times \langle \bar{\alpha}_{f},\mathbf{\bar{z}}_{f};t|\alpha _{i},\mathbf{z%
}_{i};0\rangle \rho _{\text{r}}(\mathbf{\bar{z}}_{i},\mathbf{z}%
_{i}^{\prime};0)\langle \bar{\alpha}_{i}^{\prime },\mathbf{\bar{z}}%
_{i}^{\prime};0|\alpha _{f}^{\prime },\mathbf{z}_{f};t\rangle .&
\label{effp}
\end{eqnarray}
Eq. (\ref{effp}) contains two propagators of the total system: the forward
and backward propagators, which can be expressed as path integrals. To
evaluate the forward propagator operator $e^{-i\hat{H}t}$ between the
initial ($|\alpha _{i},\mathbf{z}_{i}\rangle $) and the final ($\langle \bar{%
\alpha}_{f},\mathbf{\bar{z}}_{f}|$) coherent states, one generally divides
the time interval $t_{f}-t_{i}$ into $N$ subintervals. Then by inserting the
resolution of identity, i.e. $\int d\mu \left( \alpha \right) d\mu (\mathbf{z%
})|\alpha ,{\mathbf{z}}\rangle \langle {\bar\alpha} ,{\mathbf{\bar z}}|=1$ with the
integration measures $d\mu \left( \alpha \right) =e^{-\bar{\alpha}\alpha }%
\frac{d\bar{\alpha}d\alpha }{2\pi i}$ and $d\mu ({\mathbf{z}})=\prod_{k}e^{-%
\bar{z}_{k}z_{k}}\frac{d\bar{z}_{k}dz_{k}}{2\pi i}$, ($N-1$) times between
each subintervals and taking the limit of large $N$, the path integral
representation of the propagator can be obtained
\begin{eqnarray}
\langle \bar{\alpha}_{f},\mathbf{\bar{z}}_{f};t|\alpha _{i},\mathbf{z}%
_{i};0\rangle &=&\int D^{2}\mathbf{z}D^{2}\alpha \exp (iS_{\text{s}}[\bar{%
\alpha},\alpha ]  \notag \\
&&+iS_{\text{i}}[\mathbf{\bar{z}},\mathbf{z},\bar{\alpha},\alpha ]+iS_{\text{r}%
}[\mathbf{\bar{z}},\mathbf{z}]),  \label{prop}
\end{eqnarray}%
where
\begin{eqnarray}
&&iS_{\text{s}}[\bar{\alpha},\alpha ]={\frac{\bar{\alpha}(t)\alpha
\left(t\right) +\bar{\alpha}(0)\alpha \left( 0\right) }{2}}  \notag \\
&&~~-\int_{0}^{t}d\tau \lbrack {\frac{\bar{\alpha}(\tau )\dot{\alpha}(\tau )-%
\dot{\bar{\alpha}}(\tau )\alpha (\tau )}{2}}+iH_{\text{s}}(\bar{\alpha}%
,\alpha )], \\
&&iS_{\text{r}}[\mathbf{\bar{z}},\mathbf{z}]=\sum_{k}\{\frac{\bar{z}%
_{k}(t)z_{k}(t)+\bar{z}_{k}(0)z_{k}(0)}{2}  \notag \\
&&~~-\int_{0}^{t}d\tau \lbrack \frac{\bar{z}_{k}(\tau )\dot{z}_{k}(\tau )-%
\dot{\bar{z}}_{k}(\tau )z_{k}(\tau )}{2}+iH_{\text{r}}(\mathbf{\bar{z}},%
\mathbf{z})], \\
&&iS_{\text{i}}[\mathbf{\bar{z}},\mathbf{z},\bar{\alpha},\alpha]=-i%
\int_{0}^{t}d\tau H_{\text{i}}(\bar{\alpha},\alpha ,\mathbf{\bar{z}},\mathbf{%
z}).  \label{act}
\end{eqnarray}
are the (complex) actions corresponding to the system, reservoir, and
interaction Hamiltonian $\hat{H}_{\text{s}}$, $\hat{H}_{\text{r}}$, and $%
\hat{H}_{\text{i}}$, respectively \cite{An2009}. All the functional
integrations are evaluated over paths $\mathbf{\bar{z}}(\tau )$, $\mathbf{z}%
(\tau )$, $\bar{\alpha}(\tau )$, and $\alpha (\tau )$ with endpoints $%
\mathbf{\bar{z}}(t)=\mathbf{\bar{z}}_{f}$, $\mathbf{z}(0)=\mathbf{z}_{i}$, $%
\bar{\alpha}(t)=\alpha _{f}$, and $\alpha (0)=\alpha _{i}$. Substituting Eq.
(\ref{prop}) and a similar expression for the backward propagator into Eq. (%
\ref{effp}), we obtain
\begin{eqnarray}
&&\mathcal{J}(\bar{\alpha}_{f},\alpha _{f}^{\prime };t|\bar{\alpha}%
_{i}^{\prime },\alpha _{i};0)=\int D^{2}\alpha D^{2}\alpha ^{\prime
}\exp\{i(S_{\text{s}}[\bar{\alpha},\alpha ]  \notag \\
&&~~~~~~~~-S_{\text{s}}^{\ast }[\bar{\alpha}^{\prime },\alpha ^{\prime }])\}%
\mathcal{F[}\bar{\alpha},\alpha ,\bar{\alpha}^{\prime },\alpha ^{\prime }],
\label{effdj}
\end{eqnarray}
where
\begin{eqnarray}
&&\mathcal{F[}\bar{\alpha},\alpha ,\bar{\alpha}^{\prime },\alpha
^{\prime}]=\int d\mu (\mathbf{z}_{f})d\mu (\mathbf{z}_{i})d\mu (\mathbf{z}%
_{i}^{\prime})D^{2}\mathbf{z}D^{2}\mathbf{z}^{\prime }  \notag \\
&&~~~~~~\times \exp \{i(S_{\text{r}}[\mathbf{\bar{z}},\mathbf{z}]+S_{\text{i}%
}[\mathbf{\bar{z}},\mathbf{z},\bar{\alpha},\alpha ]-S_{\text{r}}^{\ast }[%
\mathbf{\bar{z}}^{\prime },\mathbf{z}^{\prime }]  \notag \\
&&~~~~~~-S_{\text{i}}^{\ast }[\mathbf{\bar{z}}^{\prime },\mathbf{z}^{\prime},%
\boldsymbol{\bar{\alpha}}^{\prime },\boldsymbol{\alpha }^{\prime }])\}\rho_{%
\text{r}}(\mathbf{\bar{z}}_{i},\mathbf{z}_{i}^{\prime };0)  \label{influ}
\end{eqnarray}%
is the influence functional containing all the environmental effects on the
system.

\subsection{Evaluation of the influence functional}

The path integral of the environmental part in Eq. (\ref{influ}) can be
evaluated by the saddle point method under the boundary conditions $%
z_{k}(0)=z_{ki}$, $\bar{z}_{k}(t)=\bar{z}_{kf}$. We have the equations of
motion for the stationary path as
\begin{eqnarray}
\dot{z}_{k}\left( \tau \right) +i\omega _{k}z_{k}\left( \tau
\right)&=&-ig_{k}^{\ast }\alpha \left( \tau \right) ,  \label{ss} \\
\dot{\bar{z}}_{k}\left( \tau \right) -i\omega _{k}\bar{z}_{k}\left(
\tau\right) &=&ig_{k}\bar{\alpha}\left( \tau \right) ,  \label{ss1}
\end{eqnarray}
where the paths of $\bar{\alpha}$, $\alpha $ are taken as external sources.
Substituting the formal solutions of Eqs. (\ref{ss}) and (\ref{ss1})
\begin{eqnarray}
&z_{k}\left( \tau \right) =z_{ki}e^{-i\omega _{k}\tau
}-ig_{k}^{\ast}\int_{0}^{\tau }dt^{\prime }e^{-i\omega _{k}(\tau -t^{\prime
})}\alpha\left( t^{\prime }\right)& \\
&\bar{z}_{k}\left( \tau \right) =\bar{z}_{kf}e^{-i\omega
_{k}(t-\tau)}-ig_{k}\int_{\tau }^{t}dt^{\prime }e^{i\omega _{k}(\tau
-t^{\prime })}\bar{\alpha}\left( t^{\prime }\right)&
\end{eqnarray}
into Eq. (\ref{prop}), one can obtain the result of forward propagator. It
is noted that the prefactor under the contribution of the stationary path in
the coherent-state path integral is equal to one and the saddle point method
to the evaluation of the environmental part here is exact. The similar
expression for the backward propagator $\langle \bar{\alpha}_{i}^{\prime },%
\mathbf{\bar{z}}_{i}^{\prime };0|\alpha _{f}^{\prime },\mathbf{z}%
_{f};t\rangle $ can also be determined. Assuming the initial state of the
reservoir is $\rho _{\text{r}}(0)=\frac{e^{-\beta \hat{H}_{\text{r}}}}{\text{%
Tr}_{\text{r}}e^{-\beta \hat{H}_{\text{r}}}}$, we have
\begin{eqnarray}
\rho _{\text{r}}(\mathbf{\bar{z}}_{i},\mathbf{z}_{i}^{\prime
};0)&=&\prod\limits_{k}(1-e^{-\beta \hbar \omega _{k}})\exp (-\frac{%
\left\vert z_{ik}^{\prime }\right\vert ^{2}+\left\vert \bar{z}%
_{ik}\right\vert ^{2}}{2}  \notag \\
&&+e^{-\beta \hbar \omega _{k}}\bar{z}_{ik}z_{ik}^{\prime }).
\end{eqnarray}
Substituting this initial state, the forward and backward propagators into
Eq. (\ref{influ}), and using the Gaussian integral identity $\int \frac{%
d^{2}z}{\pi }e^{-\gamma \bar{z}z+\delta z}f(\bar{z})=\frac{1}{\gamma }f(%
\frac{\delta }{\gamma })$ repeatedly, the influence functional can be
evaluated as
\begin{widetext}
\begin{eqnarray}
\mathcal{F}[\alpha ,\alpha ,\alpha ^{\prime },\alpha ^{\prime }] &=&\exp\{\int_{0}^{t}d\tau \int_{0}^{t}d\tau ^{\prime }[\nu (\tau -\tau ^{\prime })(\bar{\alpha}^{\prime }\left( \tau \right) -\bar{\alpha}\left( \tau \right))(\alpha \left( \tau ^{\prime }\right) -\alpha ^{\prime }\left( \tau^{\prime }\right) )+\,\mu (\tau -\tau ^{\prime })\bar{\alpha}^{\prime
}\left( \tau \right) \alpha \left( \tau ^{\prime }\right) ] \nonumber\\
&&-\int_{0}^{t}d\tau \int_{0}^{\tau }d\tau ^{\prime }[\mu (\tau -\tau^{\prime })\bar{\alpha}\left( \tau \right) \alpha \left( \tau ^{\prime}\right) +\mu ^{\ast }(\tau -\tau ^{\prime })\bar{\alpha}^{\prime }\left(\tau ^{\prime }\right) \alpha ^{\prime }\left( \tau \right) ]\},\label{inft}
\end{eqnarray}\end{widetext}where $\nu (\tau -\tau
^{\prime})=\sum_{k}\left\vert g_{k}\right\vert ^{2}\bar{n}(\omega
_{k})e^{-i\omega_{k}(\tau -\tau ^{\prime })}\ $and $\mu (\tau -\tau
^{\prime})=\sum_{k}\left\vert g_{k}\right\vert ^{2}e^{-i\omega _{k}(\tau
-\tau^{\prime })}$.

\subsection{Evaluation of the effective propagation functional}

With Eq. (\ref{inft}), Eq. (\ref{effdj}) can be recast into
\begin{widetext}\begin{eqnarray}
\mathcal{J}(\bar{\alpha}_{f},\alpha _{f}^{\prime };t|\bar{\alpha}'_{i},\alpha_{i};0)&=&\int D\mu \left( \alpha \right) D\mu \left( \alpha^{\prime }\right) \exp \{{1\over2}[\bar{\alpha}\left( t\right) \alpha \left(t\right) +\bar{\alpha}\left( 0\right) \alpha \left( 0\right) +\bar{\alpha}^{\prime }(t)\alpha ^{\prime }\left( t\right) +\bar{\alpha}^{\prime}(0)\alpha ^{\prime }\left( 0\right)]-A[\bar{\alpha},\alpha ;\bar{\alpha}^{\prime },\alpha ^{\prime }]\} \nonumber \\
A[\bar{\alpha},\alpha ;\bar{\alpha}^{\prime },\alpha ^{\prime }]&=&\int_{0}^{t}d\tau \{\frac{\bar{\alpha}(\tau )\dot{\alpha}(\tau )-\dot{\bar{\alpha}}(\tau )\alpha (\tau )}{2}+\frac{\dot{\bar{\alpha}}^{\prime}(\tau )\alpha ^{\prime }(\tau )-\bar{\alpha}^{\prime }(\tau )\dot{\alpha}^{\prime }(\tau )}{2}+i\omega _{0}[\bar{\alpha}(\tau )\alpha (\tau )-\bar{\alpha}^{\prime }(\tau )\alpha ^{\prime }(\tau )]  \notag \\
&&-\int_{0}^{t}d\tau ^{\prime }[\nu (\tau -\tau ^{\prime })(\bar{\alpha}^{\prime }(\tau )-\bar{\alpha}(\tau ))(\alpha (\tau ^{\prime })-\alpha^{\prime }(\tau ^{\prime }))+\mu(\tau -\tau ^{\prime })\bar{\alpha}^{\prime}(\tau )\alpha (\tau ^{\prime })]  \notag \\
&&+\int_{0}^{\tau }d\tau ^{\prime }[\mu(\tau -\tau ^{\prime })\bar{\alpha}(\tau )\alpha (\tau ^{\prime })+\mu^{\ast }(\tau -\tau ^{\prime })\bar{\alpha}^{\prime }\left( \tau ^{\prime }\right) \alpha ^{\prime }(\tau )]\}
\end{eqnarray}According to the main idea of saddle point method to evaluate the path integral, we can perform the path integral by expanding the effective action $A$ around the saddle point paths or the classical paths
\begin{eqnarray}
A[\bar{\alpha}_{\text{c}}+\delta \bar{\alpha},\alpha _{\text{c}}+\delta\alpha ;\bar{\alpha}_{\text{c}}^{\prime }+\delta \bar{\alpha}_{\text{c}}^{\prime },\alpha _{\text{c}}^{\prime }+\delta \alpha ^{\prime }] =A[\bar{\alpha}_{\text{c}},\alpha _{\text{c}};\bar{\alpha}_{\text{c}}^{\prime },\alpha _{\text{c}}^{\prime }]+\delta A+\delta ^{2}A.
\end{eqnarray}
By requesting $\delta A=0$, we obtain the equations of motion satisfied by the variables in the classical paths
\begin{eqnarray}
\dot{\alpha}_{\text{c}}(\tau )+i\omega _{0}\alpha _{\text{c}}(\tau)+\int_{0}^{t}d\tau ^{\prime }\nu (\tau -\tau ^{\prime })(\alpha _{\text{c}}(\tau ^{\prime })-\alpha_{\text{c}}^{\prime }(\tau ^{\prime}))+\int_{0}^{\tau }d\tau ^{\prime }\mu (\tau -\tau ^{\prime })\alpha _{\text{c}}(\tau ^{\prime }) =0,&&\label{a} \\
\dot{\bar{\alpha}}_{\text{c}}^{\prime }(\tau )-i\omega _{0}\bar{\alpha}_{\text{c}}^{\prime }(\tau )+\int_{0}^{t}d\tau ^{\prime }\nu ^{\ast }(\tau-\tau ^{\prime })(\bar{\alpha}_{\text{c}}^{\prime }(\tau ^{\prime })-\bar{\alpha}_{\text{c}}(\tau ^{\prime }))+\int_{0}^{\tau }d\tau ^{\prime }\mu^{\ast }(\tau -\tau ^{\prime })\bar{\alpha}_{\text{c}}^{\prime }(\tau^{\prime })=0,&& \label{abp}\\
\dot{\bar{\alpha}}_{\text{c}}(\tau )-i\omega _{0}\bar{\alpha}_{\text{c}}(\tau )+\int_{0}^{t}d\tau ^{\prime }\nu ^{\ast }(\tau -\tau ^{\prime })(\bar{\alpha}_{\text{c}}^{\prime }(\tau ^{\prime })-\bar{\alpha}_{\text{c}}(\tau ^{\prime }))+\int_{0}^{t}d\tau ^{\prime }\mu ^{\ast }(\tau -\tau^{\prime })\bar{\alpha}_{\text{c}}^{\prime }(\tau ^{\prime })-\int_{\tau
}^{t}d\tau ^{\prime }\mu ^{\ast }(\tau -\tau ^{\prime })\bar{\alpha}_{\text{c}}(\tau ^{\prime }) =0,&& \label{ab}\\
\dot{\alpha}_\text{c}^{\prime }(\tau )+i\omega _{0}\alpha ^{\prime }(\tau)+\int_{0}^{t}d\tau ^{\prime }\nu (\tau -\tau ^{\prime })(\alpha _{\text{c}}(\tau ^{\prime })-\alpha _{\text{c}}^{\prime }(\tau ^{\prime}))+\int_{0}^{t}d\tau ^{\prime }\mu(\tau -\tau ^{\prime })\alpha _{\text{c}}(\tau ^{\prime })-\int_{\tau }^{t}d\tau ^{\prime }\mu (\tau -\tau ^{\prime})\alpha _{\text{c}}\left( \tau ^{\prime }\right) =0.&&\label{app1}
\end{eqnarray}\end{widetext}We also find that the second-order term $%
\delta^{2}A$ is independent on the variables in the classical paths, which
means that it contributes only a time-dependent constant to the path
integral. Our final task is to evaluate $A[\bar{\alpha}_{\text{c}},\alpha _{%
\text{c}};\bar{\alpha}_{\text{c}}^{\prime },\alpha _{\text{c}}^{\prime }]$
using Eqs. (\ref{a}, \ref{abp}, \ref{ab}, \ref{app1}). It is interesting to
find that
\begin{equation}
A[\bar{\alpha}_{\text{c}},\alpha _{\text{c}};\bar{\alpha}_{\text{c}%
}^{\prime},\alpha _{\text{c}}^{\prime }]=0.
\end{equation}
Accordingly, we finally have
\begin{eqnarray}
\mathcal{J}(\bar{\alpha}_{f},\alpha _{f}^{\prime };t|\bar{\alpha}%
_{i}^{\prime },\alpha _{i};0) &=&M(t)\exp \{\frac{1}{2}[\bar{\alpha}%
_{f}\alpha \left( t\right) +\bar{\alpha}\left( 0\right) \alpha _{i}  \notag
\\
&&+\bar{\alpha}^{\prime }(t)\alpha _{f}^{\prime }+\bar{\alpha}%
_{i}^{\prime}\alpha ^{\prime }\left( 0\right) ]\},  \label{jtff}
\end{eqnarray}%
where $M(t)$ contributed from the second-order term can be determined by
normalization.

\subsubsection{The form of $\protect\alpha(t)$ and $\protect\alpha%
^{\prime}(0)$}

Defining $\chi (\tau )=\alpha (\tau )-\alpha ^{\prime }(\tau )\equiv
u_{1}(\tau )\chi (t)$ with $u_{1}(t)=1$, we can obtain from Eqs. (\ref{a})
and (\ref{app1})%
\begin{equation}
\dot{u}_{1}(\tau )+i\omega _{0}u_{1}(\tau )-\int_{\tau }^{t}d\tau ^{\prime
}\mu(\tau -\tau ^{\prime })u_{1}(\tau ^{\prime })=0.  \label{u1}
\end{equation}%
We can see from Eq. (\ref{a}) that the initial condition of $\alpha (\tau )$
can be fixed by $\alpha (\tau )=u(\tau )\alpha _{i}-v(\tau
)[\alpha(t)-\alpha _{f}^{\prime }]$ with $u(0)=1$ and $v(0)=0$, and
\begin{eqnarray}
\dot{u}(\tau )+i\omega _{0}u(\tau )+\int_{0}^{\tau }\mu (\tau -\tau
^{\prime})u(\tau ^{\prime })d\tau ^{\prime }=0, &&  \label{u2} \\
\dot{v}(\tau )+i\omega _{0}v(\tau )-\int_{0}^{t}\nu (\tau -\tau
^{\prime})u_{1}(\tau ^{\prime })d\tau ^{\prime } &&  \notag \\
+\int_{0}^{\tau }\mu (\tau -\tau ^{\prime })v(\tau ^{\prime })d\tau
^{\prime}=0. &&  \label{v}
\end{eqnarray}%
Comparing Eq. (\ref{u2}) with Eq. (\ref{u1}), we have $u_{1}(\tau )=u^{\ast
}(t-\tau )$. Thus we have $\alpha ^{\prime }(\tau )=u(\tau )\alpha
_{i}-v(\tau )[\alpha (t)-\alpha _{f}^{\prime }]-u_{1}(\tau )[\alpha
(t)-\alpha _{f}^{\prime }]$, which induces
\begin{equation}
\alpha ^{\prime }(0)=\alpha _{i}-u^{\ast }(t)[\alpha (t)-\alpha _{f}^{\prime
}].  \label{alppz}
\end{equation}%
Similarly, $\alpha (t)=u(t)\alpha _{i}-v(t)[\alpha (t)-\alpha _{f}^{\prime }]
$, which induces%
\begin{equation}
\alpha (t)=\frac{u(t)\alpha _{i}+v(t)\alpha _{f}^{\prime }}{1+v(t)}.
\label{alpt}
\end{equation}%
Substituing Eq. (\ref{alpt}) into Eq. (\ref{alppz}), we get
\begin{equation}
\alpha ^{\prime }(0)=[1-\frac{\left\vert u(t)\right\vert ^{2}}{1+v(t)}%
]\alpha _{i}+\frac{u^{\ast }(t)}{1+v(t)}\alpha _{f}^{\prime }.  \label{allpt}
\end{equation}

\subsubsection{The form of $\bar{\protect\alpha}(0)$ and $\bar{\protect\alpha%
}^{\prime }(t)$}

Defining $\varsigma (\tau )=\bar{\alpha}^{\prime }(\tau )-\bar{\alpha}(\tau
)\equiv u_{1}^{\prime }(\tau )\varsigma (t)$ with $u_{1}^{\prime }(t)=1$, we
can obtain from Eqs. (\ref{abp}) and (\ref{ab})%
\begin{equation}
\dot{u}_{1}^{\prime }(\tau )-i\omega _{0}u_{1}^{\prime }(\tau )-\int_{\tau
}^{t}d\tau ^{\prime }\mu^{\ast }(\tau -\tau ^{\prime })u_{1}^{\prime }(\tau
^{\prime })=0.
\end{equation}%
One can see that $u_{1}^{\prime }(\tau )=u_{1}^{\ast }(\tau )=u(t-\tau )$.
We can see from Eq. (\ref{abp}) that the initial condition of $\bar{\alpha}%
^{\prime }(\tau )$ can be fixed by $\bar{\alpha}^{\prime }(\tau )=u^{\prime
}(\tau )\bar{\alpha}_{i}^{\prime }-v^{\prime }(\tau )[\bar{\alpha}^{\prime
}(t)-\bar{\alpha}_{f}]$ with $u^{\prime }(0)=1$ and $v^{\prime }(0)=0$, and
\begin{eqnarray}
\dot{u}^{\prime }(\tau )-i\omega _{0}u^{\prime }(\tau )+\int_{0}^{\tau
}\mu^{\ast }(\tau -\tau ^{\prime })u^{\prime }(\tau ^{\prime })d\tau
^{\prime }=0, &&  \label{upm} \\
\dot{v}^{\prime }(\tau )-i\omega _{0}v^{\prime }(\tau )-\int_{0}^{t}\nu
^{\ast }(\tau -\tau ^{\prime })v^{\prime }(\tau ^{\prime })d\tau ^{\prime }
&&  \notag \\
+\int_{0}^{\tau }\mu^{\ast }(\tau -\tau ^{\prime })v^{\prime }(\tau ^{\prime
})d\tau ^{\prime }=0. &&  \label{vpm}
\end{eqnarray}%
Comparing Eqs. (\ref{upm}) and (\ref{vpm}) with Eqs. (\ref{u2}) and (\ref{v}%
), we have $u^{\prime }(\tau )=u^{\ast }(\tau )$, $v^{\prime }(\tau
)=v^{\ast }(\tau )$. So $\bar{\alpha}^{\prime }(\tau )=u^{\ast }(\tau )\bar{%
\alpha}_{i}^{\prime }-v^{\ast }(\tau )[\bar{\alpha}^{\prime }(t)-\bar{\alpha}%
_{f}]$, which induces
\begin{equation}
\bar{\alpha}^{\prime }(t)=\frac{u^{\ast }(t)\bar{\alpha}_{i}^{\prime
}+v^{\ast }(t)\bar{\alpha}_{f}}{1+v^{\ast }(t)}.  \label{aptt}
\end{equation}%
Similarly, $\bar{\alpha}(\tau )=u^{\ast }(\tau )\bar{\alpha}_{i}^{\prime
}-[v^{\ast }(t)+u(t-\tau )][\bar{\alpha}^{\prime }(t)-\bar{\alpha}_{f}]$,
which induces%
\begin{equation}
\bar{\alpha}(0)=[1-\frac{\left\vert u(t)\right\vert ^{2}}{1+v^{\ast }(t)}]%
\bar{\alpha}_{i}^{\prime }+\frac{u(t)}{1+v^{\ast }(t)}\bar{\alpha}_{f}.
\label{apto}
\end{equation}

\bigskip Substituting Eqs. (\ref{alpt}, \ref{allpt}, \ref{aptt}, \ref{apto})
into Eq. (\ref{jtff}), we get
\begin{eqnarray}
&&\mathcal{J}(\bar{\alpha}_{f},\alpha _{f}^{\prime };t|\bar{\alpha}%
_{i}^{\prime },\alpha _{i};0)=M(t)\exp [J_{1}(t)\bar{\alpha}_{f}\alpha _{i}
\notag \\
&&~~~~~-J_{2}\left( t\right) \bar{\alpha}_{f}\alpha _{f}^{\prime
}-J_{3}\left( t\right) \bar{\alpha}_{i}^{\prime }\alpha _{i}+J_{1}^{\ast }(t)%
\bar{\alpha}_{i}^{\prime }\alpha _{f}^{\prime }],  \label{prord}
\end{eqnarray}%
where
\begin{eqnarray}
M(t) &=&{\frac{1}{1+v(t)}},~J_{1}(t)={\frac{u(t)}{1+v(t)}}, \\
J_{2}(t) &=&{\frac{-v(t)}{1+v(t)}},~J_{3}(t)={\frac{|u(t)|^{2}}{1+v(t)}}-1.
\end{eqnarray}

\subsection{Exact master equation}

Eq. (\ref{timeevo}) with the final form of the effective propogation
functional given by Eq. (\ref{prord}) is the exact solution of the
decoherence dynamics of the system. On one hand, the evolution of any
initial state can be calculated by executing the integration in Eq. (\ref%
{timeevo}). On the other hand, an exact master equation can be obtained by
making time derivative to Eq. (\ref{timeevo}). In the following, we derive
the master equation. First, we have the following functional identities from
Eq. (\ref{prord})
\begin{eqnarray}
\alpha _{i}\mathcal{J} &=&\frac{1}{J_{1}\left( t\right) }[\frac{\partial }{%
\partial \bar{\alpha}_{f}}+J_{2}\left( t\right) \alpha _{f}^{\prime }]%
\mathcal{J},  \label{aaii} \\
\bar{\alpha}_{i}^{\prime }\mathcal{J} &=&\frac{1}{J_{1}^{\ast }(t)}[\frac{%
\partial }{\partial \alpha _{f}^{\prime }}+J_{2}\left( t\right) \bar{\alpha}%
_{f}]\mathcal{J},  \label{apii}
\end{eqnarray}%
which are useful to eliminating the dependence of the time derivative of the
effective propagation functional on the initial variable $\alpha _{i}$ and $%
\bar{\alpha}_{i}^{\prime }$. Then making the time derivative to Eq. (\ref%
{prord}) and the substitution of Eqs. (\ref{aaii}, \ref{apii}) and
remembering Eq. (\ref{timeevo}), we obtain the evolution equation
\begin{widetext}
\begin{eqnarray}
\dot{\rho}(\bar{\alpha}_{f},\alpha _{f}^{\prime };t) &=&\Big\{-\Gamma
^{\beta }\left( t\right) -\left[ \Gamma ^{\beta }\left( t\right) +\Gamma
\left( t\right) +i\Omega \left( t\right) \right] \bar{\alpha}_{f}\frac{%
\partial }{\partial \bar{\alpha}_{f}}+\Gamma ^{\beta }\left( t\right) \bar{%
\alpha}_{f}\alpha _{f}^{\prime }+[\Gamma ^{\beta }\left( t\right) +2\Gamma \left(
t\right) ]\frac{\partial ^{2}}{\partial \bar{\alpha}_{f}\partial \alpha
_{f}^{\prime }}  \notag \\
&&-\left[ \Gamma ^{\beta }\left( t\right) +\Gamma \left( t\right) -i\Omega
\left( t\right) \right] \alpha _{f}^{\prime }\frac{\partial }{\partial
\alpha _{f}^{\prime }}\Big\}\rho (\bar{\alpha}_{f},\alpha _{f}^{\prime };t)
\end{eqnarray}where
\begin{eqnarray}
\Gamma \left( t\right) +i\Omega \left( t\right)  =-\frac{\dot{u}(t)}{u(t)},~
\Gamma ^{\beta }\left( t\right)  =\dot{v}(t)+2v(t)\Gamma (t).\label{gammga}
\end{eqnarray}%
To obtain the operator equation of the reduced density matrix, it will be
useful to introduce the following functional differential relations. It can
be proven the following identities for arbitrary system states in the
coherent-state representation%
\begin{eqnarray}
\frac{\bar{\alpha}_{f}\partial  \rho (\bar{\alpha}_{f},\alpha _{f}^{\prime };t)}{\partial\bar{\alpha}_{f}}
\leftrightarrow \hat{a}^{\dag }\hat{a}\rho(t) ,\text{ }\bar{\alpha}_{f}\alpha
_{f}^{\prime }\rho (\bar{\alpha}_{f},\alpha _{f}^{\prime
};t)\leftrightarrow \hat{a}^{\dag }\rho(t)\hat{a},~\frac{\partial ^{2}\rho (\bar{\alpha}_{f},\alpha _{f}^{\prime };t)}{\partial
\bar{\alpha}_{f}\partial \alpha _{f}^{\prime }} \leftrightarrow \hat{a}\rho(t)
\hat{a}^{\dag },\text{ }\frac{\alpha _{f}^{\prime }\partial \rho (\bar{\alpha}_{f},\alpha _{f}^{\prime };t)}{\partial \alpha
_{f}^{\prime }}\leftrightarrow \rho(t) \hat{a}^{\dag }\hat{a},
\end{eqnarray}%
with which we arrive at our final operator form of the non-Markovian master
equation
\begin{equation}
\dot{\rho}(t)=-i\Omega \left( t\right) [\hat{a}^{\dag }\hat{a},\rho(t) ]+\Gamma \left(
t\right) \left[ 2\hat{a}\rho (t)\hat{a}^{\dag }-\hat{a}^{\dag }\hat{a}\rho (t)-\rho (t)\hat{a}^{\dag }\hat{a}%
\right] +\Gamma ^{\beta }\left( t\right) \left[ \hat{a}^{\dag }\rho (t)\hat{a}+\hat{a}\rho
(t)\hat{a}^{\dag }-\hat{a}^{\dag }\hat{a}\rho (t)-\rho (t)\hat{a}\hat{a}^{\dag }\right] .
\end{equation}\end{widetext}

\subsection{The evolved state of the coherent state as the initial state}

Consider explicitly that the system is initially in a coherent state $%
e^{-|\alpha_0|^2/2}|\alpha_0\rangle$, we have
\begin{equation}
\rho ( \bar{\alpha}_{i},\alpha _{i}^{\prime };0) =\exp ( -\left\vert
\alpha_{0}\right\vert ^{2}+\bar{\alpha}_{i}\alpha _{0}+\bar{\alpha}%
_{0}\alpha_{i}^{\prime }).  \label{bbct}
\end{equation}%
Substituting (\ref{bbct}) into Eq. (\ref{prord}) and performing the Gaussian
integration, we can get
\begin{eqnarray}
\rho ( \bar{\alpha}_{f},\alpha _{f}^{\prime };t)&=&M(t)\exp [ -(1+J_{3}(
t))\left\vert \alpha _{0}\right\vert^{2}+J_{1}(t)\bar{\alpha}_{f}\alpha _{0}
\notag \\
&&-J_{2}( t) \bar{\alpha}_{f}\alpha _{f}^{\prime}+J_{1}^{\ast }(t)\bar{\alpha%
}_{0}\alpha _{f}^{\prime }].  \label{arho}
\end{eqnarray}
Remembering that the original density matrix is expressed as $\rho(t)=\int
d\mu(\alpha_f)d\mu(\alpha_f^{\prime })\rho ( \bar{\alpha}_{f},\alpha
_{f}^{\prime };t)|\alpha_f\rangle\langle\bar{\alpha}_f^{\prime }|$, we can
readily calculate the expectation value of the bosonic number operator as
\begin{equation}
N(t)\equiv\text{Tr}_\text{s}[\hat{a}^\dag\hat{a}\rho(t)]=|%
\alpha_0u(t)|^2+v(t).  \label{an}
\end{equation}

\section{The perturbative solutions of $u(t)$ and $v(t)$}

\label{appd} In this section, we give the proof of the recovery of exact
master equation to the conventional Markovian master equation under the
second-order perturbation to the coupling strength between the system and
the reservoir.

The solution of Eq. (\ref{u2}) can be expanded as the order of the coupling
strength labeled by $\lambda $,
\begin{equation}
u(t)=u^{(0)}(t)+\lambda ^{2}u^{(2)}(t)+O(|g_{k}|^{4}),  \label{puet}
\end{equation}%
substituting which into Eq. (\ref{u2}) we can get $u^{(0)}(t)=e^{-i\omega
_{0}t}$ and
\begin{equation*}
\dot{u}^{(2)}(t)=-i\omega _{0}u^{(2)}(t)-\int_{0}^{t}\mu
(t-t_{1})u^{(0)}(t_{1})dt_{1}.
\end{equation*}%
If the characteristic time scale of the environmental correlation $\mu
(t-t_{1})$ is much smaller than the typical time scale $t$ of the system,
then we can extend the upper limit of the integration into $+\infty $. Using
the identity $\int_{0}^{+\infty }dt_{1}e^{-i(\omega -\omega _{0})t_{1}}=\pi
\delta (\omega -\omega _{0})+i{\frac{\mathcal{P}}{\omega -\omega _{0}}}$, we
get $u^{(2)}(t)=-(\kappa +i\Delta \omega )te^{-i\omega _{0}t}$, where $%
\kappa =\pi J(\omega _{0})$ and $\Delta \omega =\mathcal{P}\int {\frac{%
J(\omega )}{\omega -\omega _{0}}}d\omega $. Going back to Eq. (\ref{puet}),
we get
\begin{equation}
u(t)=e^{-(\kappa +i\omega ^{\prime })t},~\omega ^{\prime }=\omega
_{0}+\Delta \omega ,  \label{ptu}
\end{equation}%
where $\lambda $ has been set back to one.

The second-order approximate solution of Eq. (\ref{v}) reads
\begin{eqnarray}
v(t)&=&\int_{0}^{t}dt_{1}\int_{0}^{t}dt_{2}u^{*(0)}(t_{1})%
\nu(t_{1}-t_{2})u^{(0)}(t_{2})  \notag \\
&=&\int_{0}^{t}dt_{1}\int_{0}^{t}dt_{2}e^{-i(\omega-\omega_0)(t_1-t_2)}%
\nu(t_{1}-t_{2}).
\end{eqnarray}%
Since we are only concerned about its time derivative, so
\begin{eqnarray}
\dot{v}(t)&=&\int_0^\infty d\omega\int_0^td t_1
[e^{-i(\omega-\omega_0)(t-t_1)}+e^{i(\omega-\omega_0)(t-t_1)}]  \notag \\
&&\times J(\omega)\bar{n}(\omega)\simeq\kappa \bar{n}(\omega_0),  \label{ptv}
\end{eqnarray}%
where we have also used the extension of the upper limit of the integration
to $+\infty$. Substituting Eqs. (\ref{ptu}, \ref{ptv}) into Eq. (\ref{gammga}%
), we obtain the second-order perturbative values of of the parameters in
the master equation as
\begin{equation}
\Gamma(t)=\kappa,~\Omega(t)=\omega^{\prime \beta}(t)=2\kappa \bar{n}%
(\omega_0),
\end{equation}%
where the second term of $\Gamma^\beta(t)$ has been neglected because it is
the high-order term. These values are just the parameters appearing in the
master equation derived under the conventional Markovian approximation.

\section{Energy spectrum in single-excitation subspace}

Here we want to further emphasize that the roots of Eq. (8) in the main text
give the energy spectrum of the total system in the single-excitation
subspace. Therefore, we call the state represented by the discrete root $E$
as the stationary state of the total system. Furthermore, since the
single-excitation process in our Hamiltonian is solely contributed by the
vacuum of the environment, this gives further evidence on calling $u(t)$ as
the vacuum-induced dissipation function. The eigensolution in the
single-excitation subspace can be obtained in the following. The total
excitation number of our model, i.e., $\hat{N}_\text{tot}=\hat{a}^\dag\hat{a}%
+\sum_k\hat{b}_k^\dag\hat{b}$ is conserved, which means that the Hilbert
space is divided into independent subspaces with definite $\langle \hat{%
\mathcal{N}}\rangle$. The eigenstate of the total system in the subspace
with $\langle \hat{N}_\text{tot}\rangle=1$ can be expanded as%
\begin{equation}
|\varphi _{1}\rangle =c_{0}|1,\{0_{k}\}\rangle +\sum_{k}d_{k}|0,1_{k}\rangle
,
\end{equation}%
substituting which into the eigenequation, we get%
\begin{eqnarray}
c_{0}\omega _{0}+\sum_{k}g_{k}d_{k} &=&E_{1}c_{0},  \label{c} \\
c_{0}g_{k}+d_{k}\omega _{k} &=&E_{1}d_{k}.  \label{d}
\end{eqnarray}%
From Eq. (\ref{d}), we have $d_{k}=\frac{g_{k}c_{0}}{E_{1}-\omega _{k}}$,
substituting which into Eq. (\ref{c}), we obtain
\begin{equation}
\omega _{0}+\sum_{k}\frac{\left\vert g_{k}\right\vert ^{2}}{E_{1}-\omega _{k}%
}=E_{1}.
\end{equation}%
It matches well with Eq. (8) in the main text in the continuous limit of the
environmental frequency. The formation of the stationary state is in a same
mechanism as the spin-boson model case in Ref. \cite{An2010,Liu2013}.

The formation of this stationary state would cause the transient negative of the decay rate, which reveals that the system has a strong non-Markovian effect. Such effect can be quantified by the non-Markovianity \cite{Breuer}. This quantitative characterization in turn can supplies some insight to the profound impact of the formed stationary state on the dynamics of the system. The non-Markovianity is defined as
\begin{equation}
{\mathcal N}=\max_{\rho_{1}(0),\rho_2(0)}\int_{\sigma>0}\sigma(t,\rho_{1,2}(t))dt,\label{nmmkvd}
\end{equation}where $\sigma(t,\rho_{1,2}(t))={d\over dt}D(\rho_1(t),\rho_2(t))$ is the change rate of the trace distance $D(\rho_1(t),\rho_2(t))$ between two density matrix $\rho_1(t)$ and $\rho_2(t)$. The trace distance is defined as $D(\rho_1,\rho_2)=\text{Tr}|\rho_1-\rho_2|$ with $|\rho|\equiv \sqrt{\rho^\dag\rho}$. The integration is performed over all the time duration with $\sigma(t,\rho_{1,2}(t))>0$ and the maximization is performed over all the initial pair states $\rho_{1,2}(0)$. It has been proven in Ref. \cite{Breuer} that for zero-temperature environment and the one-excitation-number involved at most in the dynamics of the whole system, $\mathcal{N}$ takes the maximal value when $\rho_1(0)=|0\rangle\langle 0|$ and $\rho_2(0)=|1\rangle\langle 1|$.
\begin{figure}[tbp]
  \centering
  \includegraphics[width=0.9\columnwidth]{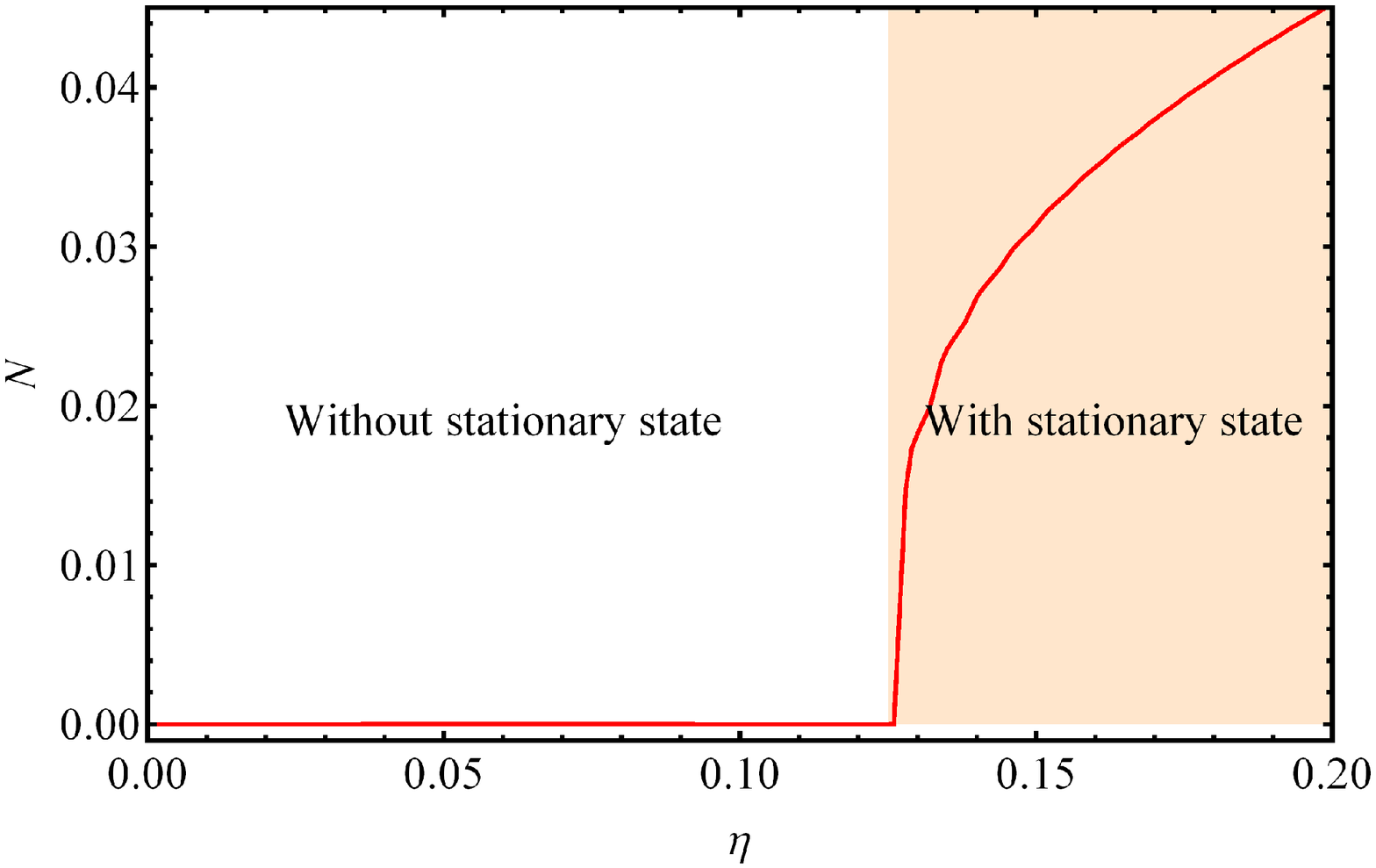}
  \caption{The behavior of the non-Markovianity $\mathcal{N}$ with the change of $\eta$ in the case of the Ohmic spectral density. $\omega_c=8.0\omega_0$ and $\beta=\infty$ have been used. The stationary state is present when $\eta>0.125$. The integration region in Eq. (\ref{nmmkvd}) is chosen from $0$ to $1600/\omega_0$, which guarantees to include all the time region with $\sigma(t,\rho_{1,2}(t))>0$.}\label{nmkv}
\end{figure}

In Fig. \ref{nmkv}, we plot the non-Markovianity calculated in the zero-temperature case, where the effect of the formed stationary state can be manifested clearly. We can see that when the stationary state is absent for the parameter regime $\eta<0.125$, $\mathcal{N}$ equals to zero persistently. Whenever the stationary state is formed, i.e., $\eta>0.125$, $\mathcal{N}$ takes non-zero value. It is understandable based on the fact the presence of the stationary state would cause the decay coefficients negative. The negative decay implies that the energy or information flows back from the environment to the system, which results in the increase of the trace distance between the two initial states.

\section{The derivation of Eq. (6)}

Eq. (6) is valid in the parameter regime without the stationary state, where
both $\Gamma(t)>0$ and $\Gamma^\beta(t)>0$ are satisfied in the whole time
evolution. In this section, we derive it analytically using the algebraic
dynamics method \cite{Wang}.

The analytical solution of Eq. (3) can be obtained by the following method.
By constructing composite algebra $\hat{N}=\hat{n}^{r}-\hat{n}^{l}$ and $\{%
\hat{K}_{0}={\frac{\hat{n}^{r}-\hat{n}^{l}}{2}},\hat{K}_{+}=\hat{a}^{r\dag }%
\hat{a}^{l},\hat{K}_{-}=\hat{a}^{r}\hat{a}^{l\dag }\}$, where $\hat{\bullet}%
^{r}$ and $\hat{\bullet}^{l}$ operates, respectively, on the right and left
vector, we can transform Eq. (3) into a Schr\"{o}dinger-equation-like form
\begin{eqnarray}
\dot{\rho}(t) &=&\mathcal{L}(t)\rho (t),  \label{ms} \\
\mathcal{L}(t) &=&-i\Omega (t)(\hat{N}+1)+A(t)\hat{K}_{+}+B(t)\hat{K}_{-}
\notag \\
&&-[A(t)+B(t)]\hat{K}_{0}-{\frac{A(t)-B(t)}{2}},
\end{eqnarray}%
where $A(t)=2\times {\frac{\Gamma ^{\beta }(t)}{2}}$ and $B(t)=2\times
\lbrack \Gamma (t)+{\frac{\Gamma ^{\beta }(t)}{2}}]$. The actions of these
operators on the state are
\begin{eqnarray}
\hat{N}|m\rangle \langle n| &=&(m-n-1)|n\rangle \langle n|, \\
\hat{K}_{0}|m\rangle \langle n| &=&{\frac{m+n+1}{2}}|m\rangle \langle n|, \\
\hat{K}_{+}|m\rangle \langle n| &=&\sqrt{(m+1)(n+1)}|m+1\rangle \langle n+1|,
\\
\hat{K}_{-}|m\rangle \langle n| &=&\sqrt{mn}|m-1\rangle \langle n-1|.
\end{eqnarray}%
$\hat{N}$ forms a u(1) algebra and $\{\hat{K}_{0}={\frac{\hat{n}^{r}-\hat{n}%
^{l}}{2}},\hat{K}_{+}=\hat{a}^{r\dag }\hat{a}^{l},\hat{K}_{-}=\hat{a}^{r}%
\hat{a}^{l\dag }\}$ form a su(1,1) algebra,
\begin{eqnarray}
\lbrack \hat{K}_{0},\hat{K}_{\pm }] &=&\pm \hat{K}_{\pm },~[\hat{K}_{-},\hat{%
K}_{+}]=2\hat{K}_{0}, \\
\lbrack \hat{N},\hat{K}_{l}] &=&0,~(l=\pm ,0).
\end{eqnarray}

With the dynamical symmetry structure at hands and by introducing a
time-dependent similarity transformation $\hat{U}=e^{\alpha_+(t)\hat{K}%
_+}e^{\alpha_-(t)\hat{K}_-}$ with $\alpha_\pm(t)$ satisfying
\begin{eqnarray}
{\frac{d\alpha_+(t)}{dt}}&=&B(t)[\alpha_+(t)-{\frac{A(t)}{B(t)}}%
][\alpha_+(t)-1],  \label{ap} \\
{\frac{d\alpha_-(t)}{dt}}&=&B(t)[1-2\alpha_+(t)\alpha_-(t)]  \notag \\
&&+[A(t)+B(t)]\alpha_-(t),  \label{alptn}
\end{eqnarray}
under the initial condition $\alpha_\pm(0)=0$, the time-dependent master
equation (\ref{ms}) can be diagonalized into
\begin{eqnarray}
\dot{\bar{\rho}}(t)&=&\hat{\bar{\mathcal{L}}}(t)\bar{\rho}(t), \\
\hat{\bar{\mathcal{L}}}(t)&=&\hat{U}^{-1}(t)\hat{\mathcal{L}}(t)\hat{U}(t)-%
\hat{U}^{-1}{\frac{d\hat{U}}{dt}}  \notag \\
&=&-i\Omega(t)(\hat{N}+1)+ \gamma(t)\hat{K}_0-{\frac{A(t)-B(t)}{2}}
\end{eqnarray}
with $\gamma(t)=2B(t)\alpha_+(t)-[A(t)+B(t)]$ and $\bar{\rho}(t)=\hat{U}%
^{-1}\rho(t)$. Then the solution of $\bar{\rho}(t)$ as well as $\rho(t)$ can
be obtained as
\begin{eqnarray}
\rho(t)&=&\hat{U}e^{\int_0^t\bar{\mathcal{L}}(\tau)d\tau}\rho(0)  \notag \\
&=&e^{\int_0^t{\frac{\gamma(\tau)-[A(\tau)-B(\tau)]}{2}} d\tau}e^{\alpha_+(t)%
\hat{K}_+}e^{\alpha_-(t)\hat{K}_-}\sum_{n,k}\rho_{n,k}  \notag \\
&&\times e^{\int_0^t[{\frac{(n+k)\gamma(\tau)}{2}}-i(n-k)\Omega(\tau)]d%
\tau}|n\rangle\langle k|,  \label{rhot}
\end{eqnarray}%
where the initial state $\rho(0)=\sum_{n,k}\rho_{n,k}|n\rangle\langle k|$
has been used.

The asymptotic form of the obtained solution (\ref{rhot}) can be analyzed by
the following method. We study first the asymptotic forms of $\alpha_\pm(t)$%
. From Eq. (\ref{ap}), we have
\begin{equation}  \label{app}
\left\{ \begin{aligned} {d\alpha_+(t)\over dt} &>0
,~~~\text{if}~0<\alpha_+(t)<{A(t)\over B(t)} \\ {d\alpha_+(t)\over dt} &<0
,~~~\text{if}~{A(t)\over B(t)}<\alpha_+(t)<1 \end{aligned} \right.
\end{equation}
when the parameters $A(t)$ and $B(t)$ satisfy $A(t)>0$ and $B(t)>0$ in the
whole time evolution. The result in (\ref{app}) reveals that $\alpha_+(t)$
evolves asymptotically from its initial value $\alpha_+(0)=0$ to the unique
stable value
\begin{equation}
\alpha_+(\infty)={\frac{A(\infty)}{B(\infty)}},  \label{apts}
\end{equation}%
under the conditions $A(t)=2\times{\frac{\Gamma^\beta(t)}{2}}>0$ and $B(t)=2%
\times[\Gamma(t)+{\frac{\Gamma^\beta(t)}{2}}]>0$, which is always satisfied
when the stationary state is absent [see Fig. 1(a,b), Fig. 2(a,b), and Fig.
6(a,b)]. To check the asymptotic form of $\alpha_-(t)$, we define $%
y(t)=\alpha_-(t)e^{\int_0^t\gamma(\tau)d\tau}$, whose time derivative reads
\begin{equation}
{\frac{dy(t)}{dt}}=[{\frac{d\alpha_-(t)}{dt}}+\alpha_-(t)\gamma(t)]e^{%
\int_0^t\gamma(\tau)d\tau}=B(t)e^{\int_0^t\gamma(\tau)d\tau},
\end{equation}%
where Eq. (\ref{alptn}) has been used. Because of
\begin{equation}
\gamma(\infty)=2B(\infty)\alpha_+(\infty)-[A(\infty)+B(\infty)]=-2\Gamma(%
\infty)<0,  \label{gm}
\end{equation}
we thus have
\begin{equation}
\lim_{t\rightarrow\infty}{\frac{dy(t)}{dt}}=0\Rightarrow y(\infty)=\text{%
Const.},  \label{ys}
\end{equation}%
which in turn implies
\begin{equation}
\alpha_-(\infty)=\infty.  \label{ants}
\end{equation}

With the asymptotic forms (\ref{apts}, \ref{ys}, \ref{ants}) at hands, we
are ready to evaluate the asymptotic form of Eq. (\ref{rhot}). Eq. (\ref%
{rhot}) can be expanded as
\begin{widetext}\begin{eqnarray}
\rho(t)=e^{\int_0^t{\gamma(\tau)-[A(\tau)-B(\tau)]\over 2} d\tau}e^{\alpha_+(t)\hat{K}_+}\sum_{n,k}\rho_{n,k}e^{\int_0^t[{(n+k)\gamma(\tau)\over 2}-i(n-k)\Omega(\tau)]d\tau}\sum_{m=0}^k{\alpha_-^m(t)\over m!}\sqrt{n!k!\over (n-m)!(k-m)!}|n-m\rangle\langle k-m|.
\end{eqnarray}Setting $m={n+k\over 2}-p$ and remembering the form of $y(t)$, we have
\begin{eqnarray}
\rho(t)&=&e^{\int_0^t{\gamma(\tau)-[A(\tau)-B(\tau)]\over 2} d\tau}e^{\alpha_+(t)\hat{K}_+}\sum_{n,k}\rho_{n,k}e^{-i\int_0^t(n-k)\Omega(\tau)d\tau}y(t)^{n+k\over 2}\nonumber\\&&\times\sum_{p={n-k\over 2}}^{n+k\over 2}{\alpha_-^{-p}(t)\over ({n+k\over 2}-p)!}\sqrt{n!k!\over (p+{n-k\over 2})!(p-{n-k\over 2})!}|p+{n-k\over2}\rangle\langle p-{n-k\over 2}|\nonumber
\end{eqnarray}\end{widetext}Due to the asymptotic form (\ref{ants}), only
the term $p=0$, i.e., $n=k$, in the summation over $p$ survives in the
long-time limit. Thus
\begin{eqnarray}
\lim_{t\rightarrow\infty}\rho(t)&=&e^{\alpha_+(\infty)\hat{K}%
_+}\sum_{n}\rho_{n,n}(\text{Const.})^{n}|0\rangle\langle 0|  \notag \\
&=&C e^{\alpha_+(\infty)\hat{K}_+}|0\rangle\langle 0|,  \label{rouf}
\end{eqnarray}%
where the asymptotic forms (\ref{ys}) and (\ref{gm}) have been used and $C$
can be determined by normalization. The final form of Eq. (\ref{rouf}) takes
the form
\begin{eqnarray}
\rho(\infty)&=&C\sum_{n=0}^\infty [{\frac{A(\infty)}{B(\infty)}}%
]^n|n\rangle\langle n|  \notag \\
&=&C\sum_{n=0}^\infty [{\frac{\Gamma^\beta(\infty)/(2\Gamma(\infty))}{%
1+\Gamma^\beta(\infty)/(2\Gamma(\infty))}}]^n|n\rangle\langle n|.
\end{eqnarray}%
Remembering both $\Gamma(\infty)$ and $\Gamma^\beta(\infty)$ are positive
values when the stationary state is absent, we can readily calculate the
normalization factor $C={\frac{1}{1+\Gamma^\beta(\infty)/(2\Gamma(\infty))}}$%
.

In summary, in obtaining Eq. (6), we are working in the parameter regime
where the stationary state is absent such that $\Gamma(t)>0$ and $%
\Gamma^\beta(t)>0$ in the whole evolution process. As soon as the stationary
state is formed, both $\Gamma(t)$ and $\Gamma^\beta(t)$ may transiently
takes negative values and asymptotically tends to zero. In this regime, Eq.
(6) does not work. It implies the breakdown of the canonical equilibration
in our system.

\end{document}